\documentclass{ieeeoj}
\usepackage{cite}
\usepackage{graphicx}
\usepackage{psfrag}
\usepackage{url}
\usepackage{amsmath}
\usepackage{array}
\usepackage{amssymb}
\usepackage{mathtools}
\usepackage{amsfonts}
\usepackage{graphicx}
\usepackage{epstopdf}
\usepackage{algorithm}
\usepackage{algorithmic}
\usepackage{setspace}
\usepackage{amscd}
\usepackage{mathrsfs}
\usepackage{epsfig}
\usepackage{color}
\usepackage{textcomp}
\let\labelindent\relax
\usepackage{enumitem}
\usepackage{amsthm}
\usepackage{tikz}
\usepackage{pgfplots}
\pgfplotsset{compat=1.17}
\usetikzlibrary{calc,intersections,arrows.meta}
\usepgfplotslibrary{fillbetween}

\usepackage{caption}
\usepackage{subcaption}
\usepackage{glossaries}
\usepackage[automake]{glossaries-extra}

\usepackage{gensymb}

\usepackage[symbol]{footmisc}

\def\BibTeX{{\rm B\kern-.05em{\sc i\kern-.025em b}\kern-.08em
		T\kern-.1667em\lower.7ex\hbox{E}\kern-.125emX}}
\AtBeginDocument{\definecolor{ojcolor}{cmyk}{0.93,0.59,0.15,0.02}}
\def\OJlogo{\vspace{-4pt}\hskip-4pt\includegraphics[height=18pt]{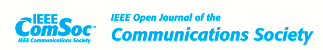}}

\newtheorem{proposition}{Proposition}

\DeclareMathOperator{\trace}{Tr}
\DeclareMathOperator*{\argmax}{argmax}

\DeclareMathOperator{\maximize}{maximize}
\DeclareMathOperator{\minimize}{minimize}

\newcommand{\Herm}[1]{{#1}^{\text{H}}}
\newcommand{\Tran}[1]{{#1}^{\text{T}}}
\newcommand{\Conj}[1]{{#1}^{\text{*}}}

\newcommand{\Brac}[1]{\left(#1\right)}
\newcommand{\Sbrac}[1]{\left[#1\right]}
\newcommand{\Cbrac}[1]{\left\{#1\right\}}

\newcommand{\Exp}[1]{\mathbb{E}\Cbrac{#1}}
\newcommand{\Abs}[1]{\left\vert #1 \right\vert}
\newcommand{\Norm}[1]{\Vert #1 \Vert}

\setabbreviationstyle[acronym]{long-short}
\newacronym{fwa}{FWA}{fixed wireless access}
\newacronym{mimo}{MIMO}{multiple-input multiple-output}
\newacronym{siso}{SISO}{single-input single-output}
\newacronym{5g}{5G}{fifth generation}
\newacronym{iid}{i.i.d.}{independent and identically distributed}
\newacronym{bs}{BS}{base station}
\newacronym{ue}{UE}{user equipment}
\newacronym{ap}{AP}{access point}
\newacronym{upa}{UPA}{uniform planar array}
\newacronym{los}{LoS}{line-of-sight}
\newacronym{awgn}{AWGN}{additive white Gaussian noise}
\newacronym{isp}{ISP}{internet service provider}
\newacronym{mmse}{MMSE}{minimum mean square error}
\newacronym{mse}{MSE}{mean square error}
\newacronym{rmse}{RMSE}{root mean square error}
\newacronym{zf}{ZF}{zero-forcing}
\newacronym{mr}{MR}{maximum-ratio}
\newacronym{mrc}{MRC}{maximum-ratio combining}
\newacronym{mrt}{MRT}{maximum-ratio transmission}
\newacronym{se}{SE}{spectral efficiency}
\newacronym{snr}{SNR}{signal-to-noise ratio}
\newacronym{sinr}{SINR}{signal-to-interference-plus-noise ratio}
\newacronym{rf}{RF}{radio frequency}
\newacronym{ofdm}{OFDM}{orthogonal frequency division multiplexing}
\newacronym{fcch}{FCCH}{frequency correction channel}
\newacronym{fb}{FB}{frequency correction burst signal}
\newacronym{gsm}{GSM}{global system for mobile}
\newacronym{crb}{CRB}{Cram{\'e}r-Rao lower bound}
\newacronym{fim}{FIM}{Fisher information matrix}
\newacronym{dft}{DFT}{discrete Fourier transform}
\newacronym{svd}{SVD}{singular value decomposition}
\newacronym{cp}{CP}{cyclic prefix}
\newacronym{lo}{LO}{local oscillator}
\newacronym{tdd}{TDD}{time division duplexing}
\newacronym{ml}{ML}{maximum likelihood}
\newacronym{csi}{CSI}{channel state information}
\newacronym{pcsi}{PCSI}{perfect channel state information}
\newacronym{cpu}{CPU}{central processing unit}
\newacronym{fgb}{FGB}{fixed grid of beams}
\newacronym{nls}{NLS}{non-linear least squares}
\newacronym{fc}{FC}{fully coherent}
\newacronym{fnc}{FNC}{fully non-coherent}
\newacronym{pcnc}{PC}{partially coherent}
\newacronym{wmmse}{WMMSE}{weighted minimum-mean-square-error}
\newacronym{bcd}{BCD}{block coordinate descent}
\newacronym{qcqp}{QCQP}{quadratically constrained quadratic program}
\newacronym{sic}{SIC}{successive interference cancellation}
\newacronym{cinr}{CINR}{channel-to-noise-plus-interference-channel ratio}

\makeglossaries

\DeclareMathOperator{\tH}{\mathrm{H}}

\DeclareMathOperator{\C}{\mathbb{C}}

\DeclareMathOperator{\K}{\mathcal{K}}

\DeclareMathOperator{\KK}{\mathbf{K}}

\DeclareMathOperator{\OO}{\mathcal{O}}

\DeclareMathOperator{\E}{\mathbf{E}}
\DeclareMathOperator{\V}{\mathbf{V}}

\DeclareMathOperator{\w}{\mathbf{w}}

\DeclareMathOperator{\HH}{\mathbf{H}}

\DeclareMathOperator{\G}{\mathbf{G}}

\DeclareMathOperator{\T}{\mathbf{T}}

\DeclareMathOperator{\LL}{\mathcal{L}}
\DeclareMathOperator{\SSS}{\mathcal{S}}
\DeclareMathOperator{\CN}{\mathcal{CN}}
\DeclareMathOperator{\A}{\mathbf{A}}

\DeclareMathOperator{\B}{\mathbf{B}}
\DeclareMathOperator{\bb}{\mathbf{b}}

\DeclareMathOperator{\CC}{\mathbf{C}}
\DeclareMathOperator{\CCC}{\mathcal{C}}
\DeclareMathOperator{\ccc}{\mathbf{c}}

\DeclareMathOperator{\DD}{\mathcal{D}}

\DeclareMathOperator{\W}{\mathbf{W}}

\DeclareMathOperator{\PP}{\mathbf{P}}

\DeclareMathOperator{\II}{\mathbf{I}}

\DeclareMathOperator{\y}{\mathbf{y}}

\DeclareMathOperator{\U}{\mathbf{U}}

\DeclareMathOperator{\q}{\mathbf{q}}

\DeclareMathOperator{\X}{\mathbf{X}}
\DeclareMathOperator{\Y}{\mathbf{Y}}

\DeclareMathOperator{\ZZ}{\mathcal{Z}}

\DeclareMathOperator{\SIGMA}{\boldsymbol{\Sigma}}

\DeclareMathOperator{\PSI}{\boldsymbol{\Psi}}

\DeclareMathOperator{\LAMBDA}{\boldsymbol{\Lambda}}

\makeatletter
\newcommand\fs@spaceruled{\def\@fs@cfont{\bfseries}\let\@fs@capt\floatc@ruled
\def\@fs@pre{\vspace{0.5\baselineskip}\hrule height.7pt depth0pt \kern2pt}%
\def\@fs@post{\kern2pt\hrule\relax}%
\def\@fs@mid{\kern2pt\hrule\kern2pt}%
\let\@fs@iftopcapt\iftrue}
\makeatother

\newcommand{\revision}[1]{#1}

\allowdisplaybreaks

\begin{document}

\receiveddate{07 February, 2024}
\accepteddate{29 February, 2024}
\doiinfo{OJCOMS.2024.3373170}

\jvol{XX}
\pubyear{202X}

\title{Cell-Free Massive MIMO with Multi-Antenna Users and Phase Misalignments: A Novel Partially Coherent Transmission Framework}

\author{UNNIKRISHNAN KUNNATH GANESAN\authorrefmark{1} (Graduate Student Member, IEEE), \\TUNG THANH VU\authorrefmark{2} (Member, IEEE), ERIK G. LARSSON\authorrefmark{1} (Fellow, IEEE)}
\affil{Department of Electrical Engineering (ISY), Linköping University, 581 83 Linköping, Sweden}
\affil{School of Engineering, Macquarie University, Australia}
\authornote{This work is supported in part by ELLIIT, in part by KAW foundation, and in part by the REINDEER project of the European Union‘s Horizon 2020 research and innovation programme under grant agreement No. 101013425.}
\markboth{A Novel Partially Coherent Transmission Framework for Cell-Free Massive MIMO}{Kunnath Ganesan \textit{et al.}}

\begin{abstract}
Cell-free massive \gls{mimo} is a promising technology for next-generation communication systems. 
This work proposes a novel \gls{pcnc} transmission framework to cope with the challenge of phase misalignment among the \glspl{ap}, which is important for unlocking the full potential of cell-free massive \gls{mimo} technology. 
With the \gls{pcnc} operation, the \glspl{ap} are only required to be phase-aligned within clusters. 
Each cluster transmits the same data stream towards each \gls{ue}, while different clusters send different data streams. 
We first propose a novel algorithm to group \glspl{ap} into clusters such that the distance between two \glspl{ap} is always smaller than a reference distance ensuring the phase alignment of these \glspl{ap}. 
Then, we propose new algorithms that optimize the combining at \glspl{ue} and precoding at \glspl{ap} to maximize the downlink sum data rates. 
We also propose a novel algorithm for data stream allocation to further improve the sum data rate of the \gls{pcnc} operation. 
Numerical results show that the \gls{pcnc} operation using the proposed framework with a sufficiently small reference distance can offer a sum rate close to the sum rate of the ideal \gls{fc} operation that requires network-wide phase alignment. 
This demonstrates the potential of  \gls{pcnc} operation in practical deployments of cell-free massive \gls{mimo} networks.   
\end{abstract}


\maketitle

\glsresetall

\section{Introduction} \label{sec:Introduction}
\IEEEPARstart{C}{ell-free} massive \gls{mimo} is an innovative technology for next-generation communication systems, where \glspl{ue} are served by a large number of \glspl{ap} distributed over an extensive geographic area~\cite{nayebi2015cell,ngo2017cell,interdonato2019ubiquitous,demir2021foundations,ngo2015cell,elhoushy2022cell,bjornson2020scalable,zhang2019cell}.
It inherits the multi-antenna benefits of cellular massive \gls{mimo}~\cite{marzetta2010noncooperative,marzetta2016fundamentals} and provides extraordinary macro diversity gains~\cite{interdonato2019ubiquitous}. 
In cell-free massive \gls{mimo} networks, inter-cell interference is eliminated due to the absence of rigid cell boundaries, providing uniformly great service to all \glspl{ue}.

Precise phase alignment of the \glspl{ap} is crucial to unlock the full potential of cell-free massive \gls{mimo}~\cite{vieira2021reciprocity,ganesan2023beamsync,larsson2024massive,nissel2022correctly}. 
With phase alignment, all the \glspl{ap} can coordinate together and transmit the same data stream toward each \gls{ue}, leading to a high beamforming/array gain, and hence, a high data rate.
In this case, the \glspl{ap} are working in a \gls{fc} operation of cell-free massive \gls{mimo} systems. 
\revision{Note that the \gls{fc} operation represents an upper bound on the performance of cell-free networks.}

However, ensuring a full phase synchronization of all the \glspl{ap} in the preferable \gls{fc} operation of cell-free massive \gls{mimo} networks is practically very challenging~\cite{vieira2021reciprocity,chen2022survey}. 
The \glspl{ap} are driven by independent \glspl{lo} that may drift independently with time, which results in phase mismatches between them.  
This requires frequent re-alignment of the phase between different \glspl{ap}~\cite{nissel2022correctly}, which is infeasible for large-scale networks. 
This raises the need for finding solutions to operate cell-free massive \gls{mimo} networks in the presence of phase alignment errors, to unlock all the benefits of these networks.
     
There are two operations to cope with the problem of phase misalignment in cell-free massive \gls{mimo} networks. 
The first is a \gls{fnc} operation, where the \glspl{ap} are not phase-aligned and send independent data streams toward \glspl{ue}~\cite{zheng2023asynchronous,vu2020noncoherent,bai2023distributed,ammar2022downlink,ozdogan2019performance} (see Section~\ref{sec:relatedwork} for more details). 
The second is a \gls{pcnc} operation, allowing both coherent and non-coherent transmissions of \glspl{ap} in the same cell-free massive \gls{mimo} network~\cite{antonioli2023mixed}. 
With \gls{pcnc} operation, there are multiple clusters, in each of which the \glspl{ap} are aligned in phase. 
The \glspl{ap} in the same cluster send the same data stream toward each \gls{ue}, while those in different clusters send different data streams. 
The \gls{pcnc} operation is more practical compared to the \gls{fc} operation as phase alignment is only performed among \glspl{ap} within a relatively small area.  
Such \gls{pcnc} operation can offer higher data rates compared to the \gls{fnc} operation.
\revision{The \gls{pcnc} operation does not require network-wide synchronization, which makes it practically feasible and attractive for realizing future cell-free massive \gls{mimo} communication systems.}

In this paper, we propose an innovative framework for the \gls{pcnc} operation in cell-free massive \gls{mimo} networks. 
First, an algorithm is proposed to group \glspl{ap} into multiple phase-aligned clusters, which is fundamentally different from the current \gls{ap} clustering algorithms where the phase misalignment is not taken into account. 
Then, novel signal processing and data stream allocation algorithms are developed to significantly improve the performance, compared to the \gls{fnc} operation. 
Our \gls{ap} clustering method is based on the insights from recent advances in phase alignment techniques proposed in~\cite{vieira2021reciprocity,ganesan2023beamsync,larsson2024massive}. 
Even though network-wide phase alignment might be infeasible, these works show that it is possible to synchronize the phases of \glspl{ap} that are close to each other in the network. 
This leads to two research questions: (Q1) What is the maximum ``reference'' distance between the \glspl{ap} such that they are considered to be in a phase-aligned cluster?
(Q2) Given this reference distance, how to cluster the \glspl{ap}, and how to improve the performance of the system?
In this work, we focus on answering the question (Q2). Answering question (Q1) requires detailed studies from field measurements, which is out of the scope of this paper.

\subsection{Related Literature and Discussions} \label{sec:relatedwork}
A large literature on the \gls{fc} operation of cell-free massive \gls{mimo} systems is available, assuming all the \glspl{ap} are operating coherently without considering the problem of phase misalignment. \revision{Comprehensive surveys of the field are provided in ~\cite{demir2021foundations,chen2022survey}.}
The advantages of cell-free massive \gls{mimo} in terms of energy and cost efficiency are considered in~\cite{zhang2019cell}. 
The survey paper~\cite{chen2022survey} provides a comprehensive study on the centralized and distributed operations of cell-free massive \gls{mimo}.
Uplink performance analysis is studied in~\cite{papazafeiropoulos2020performance} and under limited fronthaul and hardware impairments are studied in~\cite{masoumi2020performance}. 
The paper~\cite{buzzi2017cell} studies the user-centric operation of cell-free massive \gls{mimo} to reduce the fronthaul overhead. 
To maximize the uplink sum rate of \glspl{ue}, a max-min approach under power constraints is studied in~\cite{bashar2019uplink}.  
A centralized precoding and power control strategy is proposed in~\cite{kaya2020dense} when users are served by overlapping clusters of \glspl{ap}. 
The paper~\cite{bjornson2019new} studies a scalable implementation of distributed massive \gls{mimo} systems exploiting dynamic cooperation of clusters and considers initial access, pilot assignment, cooperation cluster formation, precoding, and combining strategies. 
These existing works have shown that cell-free massive \gls{mimo}  with perfect phase alignment offers significant gains in terms of spectral and energy efficiencies, compared to cellular massive \gls{mimo} networks. 
	
The \gls{fnc} operation of cell-free networks, which do not fully exploit the multi-antenna benefits of cell-free massive \gls{mimo}, have been studied in~\cite{zheng2023asynchronous,vu2020noncoherent,bai2023distributed,ammar2022downlink,ozdogan2019performance}. 
The asynchronous arrival of signals at the \glspl{ue} increases the interference and degrades the system performance. 
The work~\cite{zheng2023asynchronous} proposes a rate-splitting strategy by splitting the messages into common and private parts to improve the data rate with non-coherent operation between the \glspl{ap}.
The paper~\cite{vu2020noncoherent} investigates the non-coherent joint transmission and poses the beamforming vector design as an optimization problem, with the objective of maximizing the sum rate of the system.
This optimization problem is non-convex and NP-hard and different schemes have been proposed in the literature to achieve near-optimal solutions~\cite{vu2020noncoherent,bai2023distributed,ammar2022downlink}.
The work~\cite{vu2020noncoherent} develops an algorithm based on the alternating direction method of multipliers with an inner approximation technique to achieve a near-optimal solution. 
The paper~\cite{bai2023distributed} proposes an algorithm based on multi-agent reinforcement learning to maximize the sum rate with low computational complexity. 
The paper~\cite{ammar2022downlink} uses tools from fractional programming, block coordinate descent, and compressed sensing to construct a smooth non-decreasing algorithm to optimize the beamforming vectors.
The spectral efficiency of cell-free massive \gls{mimo} systems under Rician fading channels and phase shifts of the \gls{los} path is studied in~\cite{ozdogan2019performance}. 
These works have shown the significant importance of optimizing beamforming in improving the performance of cell-free massive \gls{mimo} networks with \gls{fnc} operation.

The research on the \gls{pcnc} operation of cell-free massive \gls{mimo} networks is still in its infancy, and we are only aware of one related paper~\cite{antonioli2023mixed}. 
The work~\cite{antonioli2023mixed} proposed a mixed coherent and non-coherent transmission scheme for cell-free systems with single-antenna \glspl{ue}. 
Here, there are multiple \glspl{cpu}, and the \glspl{ap} connect to a \gls{cpu} and operate coherently within a cluster. 
Different \glspl{cpu} or different clusters work in a non-coherent fashion. 
Reference \cite{antonioli2023mixed} showed that a mixed approach performs in between the coherent and non-coherent approaches. 
\revision{However, \cite{antonioli2023mixed} assumed that the \gls{ap} clusters are already known, and did not take into account the practical problem of phase misalignment. 
Moreover, \cite{antonioli2023mixed} considered fixed beamforming and did not consider the aspect of optimizing the beamforming to achieve maximum beamforming gains and data rates for the \gls{pcnc} operation.}

\label{R2C1}
\revision{
Clustering in cell-free massive \gls{mimo} has been studied in~\cite{demir2021foundations,bjornson2020scalable,bjornson2019new,al2021multiple,le2021learning,banerjee2023access,ammar2022downlink}. 
The purpose of the clustering methods in these papers is to maximize the spectral efficiency or energy efficiency with limited-capacity fronthaul. 
None of them takes into account the phase misalignment problem.
Moreover, the works on user clustering and cooperative transmissions in cell-free networks assume network-wide synchronization of the \glspl{ap}. 
In practice, network-wide full synchronization is not feasible. 
Using state-of-the-art methods for synchronization described in~\cite{vieira2021reciprocity,ganesan2023beamsync,larsson2024massive}, it is possible to synchronize \glspl{ap} within a small area. 
How such areas and clusters of \glspl{ap} can be determined has not been studied so far. 
In this work, we propose a \gls{pcnc} framework to overcome the challenge of phase misalignment. It consists of an \gls{ap} clustering algorithm to determine the \glspl{ap} within a reference distance that can be fully synchronized. 
}

The studies on the systems with multi-antenna \glspl{ue} have been studied primarily for cellular massive \gls{mimo}~\cite{sutton2021hardening,kazemi2023robust,bjornson2013receive}, and recently for cell-free massive \gls{mimo} in~\cite{mai2020downlink,wang2023uplink,li2022cell}.
Optimizing combining, precoding, and data stream allocation is important for these systems. 
It is normally preferred to achieve higher multiplexing gains by using multiple data streams at the same time. 
However, multiplexing more data streams towards one \gls{ue} can bring more inter-stream interference, which degrades the data rates and requires optimized combining/precoding techniques to mitigate. 
A proper method to allocate data streams can strike the best trade-off between high multiplexing gains and low inter-stream interference. 
The work~\cite{sutton2021hardening} proposes a \gls{mr} precoding scheme to enhance the channel hardening effect in systems with limited hardening capabilities in cellular massive \gls{mimo}.
A joint precoding and combiner design in the presence of reciprocity calibration errors is studied in~\cite{kazemi2023robust}.
Data stream allocation for multi-antenna \glspl{ue} for a cellular massive \gls{mimo} system is studied in~\cite{bjornson2013receive}.
Downlink \gls{se} of cell-free massive \gls{mimo} is derived in~\cite{mai2020downlink} considering imperfect \gls{csi}, non-orthogonal pilots and power control. 
An iterative \gls{wmmse} precoding scheme for uplink cell-free massive \gls{mimo} is proposed in~\cite{wang2023uplink} which shows a higher \gls{se} with a large number of \gls{ue} antennas.
An eigenbasis-based uplink precoding scheme is proposed in~\cite{li2022cell} to improve the \gls{se}. 
These existing works focus on the \gls{fc} operation and do not consider the \gls{pcnc} operation and data stream allocation for cell-free massive \gls{mimo} networks. 
	
\subsection{Research Gap and Main Contributions}\label{sec:MainContributions}

\Gls{pcnc} is an innovative and practical way of unlocking the full potential of cell-free massive \gls{mimo} networks. 
To the best of our knowledge, there is no systematic study of this operation available in the literature. 
The designs of \gls{ap} clustering based on the reference distance, precoding/combining optimization, and data stream allocation significantly impact the data rate performance of the \gls{pcnc} operation in cell-free massive \gls{mimo} systems.
By proposing an innovative framework that involves these aspects, the paper makes the following contributions:
\begin{itemize}
 
    \item We propose an \gls{ap} clustering algorithm for cell-free massive \gls{mimo} systems, providing a set of non-overlapping phase-aligned clusters for  \gls{pcnc} operation. 
    The proposed algorithm groups the \glspl{ap} into phase-aligned clusters based on both the reference distance of phase alignment and the channel conditions, which is different from that in \cite{antonioli2023mixed}.
    
    \item We develop an algorithm for optimizing the combining and precoding matrices to maximize the downlink sum rate. 
    The formulated optimization problem involves per-\gls{ap} power constraints and varying-size variable matrices. 
    It is also a non-convex and NP-hard problem, which is very challenging to solve for global optimality. 
    We propose an algorithm that tailors the \gls{wmmse} framework to obtain a sub-optimal solution to the formulated optimization problem. 
    Importantly, the developed algorithm only requires calculating closed-form expressions for the combining and precoding matrices in each iteration, which is computationally efficient for large-scale networks. 
    Note that the \gls{wmmse} algorithms proposed in existing works~\cite{shi2011iteratively,shi2015secure} cannot be directly applied to solve the formulated problem. 
    \revision{We show that our proposed scheme performs significantly better than the fixed beamforming techniques in~\cite{antonioli2023mixed}.}
    
    \item We propose a greedy data stream allocation algorithm for multi-antenna \glspl{ue}, which further maximizes the sum rate of the network. 
    The data stream of each \gls{ue} is iteratively allocated to have both strong channel gains and low interference strengths.
    
    \item We analyze numerically the performance of the \gls{pcnc} system with the proposed algorithms. 
    We also make detailed comparisons in terms of sum data rates between the \gls{pcnc} and the traditional operations \gls{fc} and \gls{fnc}.
    Numerical results show that the \gls{pcnc} operation with phase-aligned \gls{ap} clusters clustered based on an appropriate reference distance offers similar performance as that of the \gls{fc} operation. 
    This means that it is possible to approach the ideal performance of the \gls{fc} operation by using  \gls{pcnc} operation, without the strict requirement of network-wide phase alignment. 
    This highlights the importance of the \gls{pcnc} operation in the practical deployment of cell-free massive \gls{mimo} networks. 
    The results also show that data stream allocation plays an important role in improving the performance of the \gls{pcnc}.

\end{itemize}

The rest of the paper is organized as follows. 
Section~\ref{sec:ApOperation} provides a motivational example to explain why studying the operation of non-phase-aligned \glspl{ap} is  important. 
Section~\ref{sec:SystemModel} introduces the system model and problem formulation. 
The proposed \gls{ap} clustering algorithm is presented in Section~\ref{sec:PCNC_ApClustering}. 
Section~\ref{sec:ans:DesignCombPrecoder} provides algorithms to optimize precoding and combining matrices to maximize the sum rate of the network for a \gls{pcnc} system, while Section~\ref{sec:DataStreamAllocation} provides the greedy data-stream allocation and Section~\ref{sec:ComplexityAnalysis} discusses the complexity of the proposed algorithms.
Numerical results are provided in Section~\ref{sec:Results} and concluding remarks are given in Section~\ref{sec:Conclusion}.

\textit{Notations:} Bold lowercase letters are used to denote vectors and bold uppercase letters are used to denote matrices. $\mathbb{C}$ denotes the set of complex numbers. For a matrix $\mathbf{A}$, $\Conj{\mathbf{A}}$, $\Tran{\mathbf{A}}$ and $\Herm{\mathbf{A}}$ denote conjugate, transpose and conjugate transpose of the matrix $\mathbf{A}$ respectively. $\mathcal{CN}(0,\sigma^2)$ denotes a circularly symmetric complex Gaussian random variable with zero mean and variance equal to $\sigma^2$. $\text{Tr}(\mathbf{A})$ denotes the trace of $\mathbf{A}$. The identity matrix of size $ N\times N $ is denoted by $ \mathbf{I}_N $. 

\section{Motivating Example} 
\label{sec:ApOperation}
	
In this section, we consider a small example to understand the motivation of the \gls{pcnc} operation and the importance of beamforming in operating \glspl{ap} with phase misalignment.
Consider two \glspl{ap}, each with $M$ antennas, and a single \gls{ue} with $N$ antennas.
Let $\mathbf{G}_{1}$ and $\mathbf{G}_{2}$ be the $N\times M$ channel between the \glspl{ap} $1$ and $2$ to the \gls{ue}, respectively. 
We assume that the channels are known to every entity.
Let $\rho$ be the operating \gls{snr} (normalized transmit power) at each \gls{ap}.
	
\subsection{Phase Alignment Scenario} \label{sec:PhaseAlignedScenario}
When both  \glspl{ap} are phase-aligned, they can form a virtual large massive \gls{mimo} array with $2M$ elements and operate coherently together. 
This is analogous to a point-to-point \gls{mimo} system and the achievable rate is $\log_2 \Abs{ \II + \rho \HH \KK \HH^{\tH} }$~\cite[Eq. (C.28)]{marzetta2016fundamentals}, where $\HH = [\G_{1} \G_{2}]$ is the combined \gls{mimo} channel and 
\begin{align}
	\KK = \begin{bmatrix}
	\KK_{11} & \KK_{12} \\
	\KK_{21} & \KK_{22}
	\end{bmatrix}
\end{align}
is the covariance matrix of the signal transmitted from the virtual large massive \gls{mimo} array.
Each \gls{ap} has a maximum transmit power constraint given by
\begin{align}
	\trace\Brac{\KK_{11}} \leq 1 , ~~~  \trace\Brac{\KK_{22}} \leq 1.
\end{align}

The achievable rate under per-\gls{ap} power constraints is determined by the following maximization problem:
\begin{subequations}
\begin{align}
    R_{\text{aligned}}  = \underset{\KK}{\max} & ~~ \log_2 \Abs{ \II + \rho \HH \KK \HH^{\tH} } 
    \\
     \text{subject to}  & ~~ \trace\Brac{\KK_{11}} \leq 1, ~~~  \trace\Brac{\KK_{22}} \leq 1.
\end{align}
\end{subequations}
This problem can be solved to global optimality using standard convex programming software packages. 
$R_{\text{aligned}}$ will be used as the ideal data rate for comparisons with the data rates in practical scenarios of phase misalignment.
	
\subsection{Phase Misalignment Scenario}\label{sec:PhaseMisAlignedScenario}
	
When both \glspl{ap} are not phase-aligned, the relative phase between the two \glspl{ap} is unknown. 
This scenario is likely to happen in practice due to the drift of the clocks between the \glspl{ap} unless a strict phase-synchronization protocol is implemented. 
To cope with this phase misalignment problem, several transmission strategies can be used as follows.

\subsubsection{Transmit  Only from  the Best AP} 
The first strategy is to simply let \gls{ue} pick the \gls{ap} that gives the maximum rate and receives its data stream from that \gls{ap} only. 
The \gls{ue} achievable rate in this case is 
\begin{align}
    R_{\text{best AP}} = \max \{R_{\text{\gls{ap} $1$ only}}, R_{\text{\gls{ap} $2$ only}} \},
\end{align}
where $R_{\text{\gls{ap} i only}}$ is the data rate when only \gls{ap} $i, \forall i \in \{1,2\},$ is in operation is 
\begin{subequations}
\begin{align}
    R_{\text{\gls{ap} i only}} = \underset{\KK_{ii}}{\max} & ~~ \log_2 \Abs{ \II + \rho \G_{i} \KK_{ii} \G_{i}^{\tH} } 
    \\
    \text{subject to} & ~~ \trace\Brac{\KK_{ii}} \leq 1.
\end{align}    
\end{subequations}

\subsubsection{Transmit Independently Coded Data Streams with Optimal Beamforming and Successive Interference Cancellation}
Another strategy is to transmit independently coded data from the two \glspl{ap}. 
The transmitted signals from \glspl{ap} $1$~and~$2$ have the covariance matrices $\KK_{11}$ and $\KK_{22}$, respectively. 
The \gls{ue} applies \gls{sic} and the achievable rate is given by~\cite[Sec. 8.3.3]{tse2005fundamentals}
\begin{subequations}
\begin{align}
     \!\!R_{\text{SIC}} = \! \underset{\KK_{11}, \KK_{22}}{\max} & \log_2\! \Abs{ \II \!+ \rho \G_{1} \KK_{11} \G_{1}^{\tH} + \rho \G_{2} \KK_{22} \G_{2}^{\tH}} 
     \\
    \text{subject to}\,\, &\! \trace\Brac{\KK_{11}} \leq 1, ~ \trace\Brac{\KK_{22}} \leq 1.
\end{align}    
\end{subequations}
This approach can be used when we have multi-point to single-point communication. 
However, in scenarios with multiple users, this approach requires all users to know the channels of all other users, and hence, becomes infeasible in practice. 
Thus, it is necessary to look for sub-optimal strategies.

\subsubsection{Transmit Independently Coded Data Streams with Sub-optimal Beamforming}
\label{sec:subOptimalBeamforming}
Another strategy is to find sub-optimal solutions for beamforming from each \gls{ap} and combining at the \gls{ue}. 
For example, consider a simple case where \glspl{ap} beamform a rank-one signal each in the direction $\mathbf{w}_1$ and $\mathbf{w}_2$, respectively, satisfying the power constraints at each \gls{ap}. 
If a \gls{zf} combiner is used at the \gls{ue}, the rate obtained is given by 
\begin{align}
	R_{\text{non-aligned}} = \log_2\Brac{1 + \frac{\rho}{\sigma_1^2}} + \log_2\Brac{1 + \frac{\rho}{\sigma_2^2}} ,
\end{align}
where $\sigma_i^2 = \Sbrac{ \Brac{\HH^{\tH} \HH}^{-1} }_{ii}, \forall i \in \{1,2\}$, and $\HH = \Sbrac{\G_1 \w_1 ~ \G_2 \w_2} \in \mathbb{C}^{N\times 2}$ is the overall effective channel. 
Then, the remaining problem is to select the beamformers $\w_1$ and $\w_2$ to maximize the rate under the power constraints. 
A natural approach is to select the beamformers $\w_1$ and $\w_2$ to be the dominant right singular vectors of $\G_1$ and $\G_2$, respectively, to maximize the per-stream \gls{snr} from each \gls{ap}. 
The selection of beamformers must also consider the angle between $\G_1\w_1$ and $\G_2\w_2$ as well.


To gain more insights \revision{on the above concept}, we consider an example with \revision{$N=2$ such that the \gls{ue} receives signals from two directions and}
\begin{align}
	\label{eqn:analyticalExample1}
	\G_1 & = \mathbf{g} \mathbf{a}^{\tH} \\
	\label{eqn:analyticalExample2}
	\G_2 & = \mathbf{g} \mathbf{b}^{\tH}  + \alpha \mathbf{f} \mathbf{c}^{\tH}.
\end{align}
Here, $\Norm{\mathbf{g}} = 1 $, $\Norm{\mathbf{f}} = 1 $, $\Norm{\mathbf{a}} = 1 $, $\Norm{\mathbf{b}} = 1 $, $\Norm{\mathbf{c}} = 1 $ and $\Abs{\alpha} < 1$ is some constant, and such that $\mathbf{g}^{\tH}\mathbf{f}~=~0$ and $\mathbf{c}^{\tH}\mathbf{b}~=~0$. 
\revision{The above channels correspond to a situation where the signal to the \gls{ue} arrives from \gls{ap} $1$ from a single direction (direction $\mathbf{g}$) and signal from \gls{ap} $2$ arrives from two directions: (i) from direction $\mathbf{g}$ (same direction as \gls{ap}-$1$); (ii) a weaker signal from direction $\mathbf{f}$.}
Selecting dominant right singular vectors as the beamformers, i.e., $\w_1 = \mathbf{a}$ and $\w_2 = \mathbf{b}$, gives $\G_1\w_1 =\mathbf{g}$ and $\G_2\w_2 =\mathbf{g}$. 
This makes the effective channel $\mathbf{H}$ non-invertible, creating huge interference during decoding. 
\revision{Instead, choosing $\w_2 = \mathbf{c}$, the second-strongest singular vector of $\G_2$, we have $\G_1\w_1 = \mathbf{g}$ and $\G_2\w_2 = \alpha \mathbf{f}$. 
Thus the second option is better, even though the signal from \gls{ap} $2$ in the direction of $\ccc$ is weaker than that in the direction of $\bb$. }

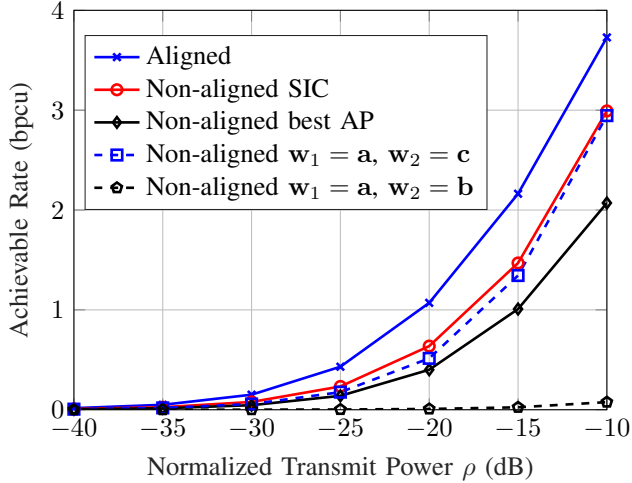
\begin{figure}[!t]
    \centering
%
%
\begin{tikzpicture}

\begin{axis}[%
width=0.4\textwidth,
height=0.3\textwidth,
at={(0.758in,0.481in)},
scale only axis,
xmin=-40,
xmax=-10,
xlabel style={font=\color{white!15!black}},
xlabel={Normalized Transmit Power $\rho$ (dB)},
ymin=0,
ymax=4,
ylabel style={font=\color{white!15!black}},
ylabel={Achievable Rate (bpcu)},
axis background/.style={fill=white},
xmajorgrids,
ymajorgrids,
legend style={at={(0.02,0.5)}, anchor=south west, legend cell align=left, align=left, draw=white!15!black}
]
\addplot [color=blue, line width=1.0pt, mark=x, mark options={solid, blue}]
  table[row sep=crcr]{%
-40	0.0157849749092841\\
-35	0.0493372367521921\\
-30	0.150577694469058\\
-25	0.43070745976863\\
-20	1.07048461653928\\
-15	2.16315861332258\\
-10	3.72843479656137\\
};
\addlegendentry{Aligned}

\addplot [color=red, line width=1.0pt, mark=o, mark options={solid, red}]
  table[row sep=crcr]{%
-40	0.00797636325457669\\
-35	0.0250748822548557\\
-30	0.0778551094477133\\
-25	0.233145063990011\\
-20	0.637180711899181\\
-15	1.46963062294562\\
-10	2.99228471609576\\
};
\addlegendentry{Non-aligned SIC}

\addplot [color=black, line width=1.0pt, mark=diamond, mark options={solid, black}]
  table[row sep=crcr]{%
-40	0.00460924984467394\\
-35	0.0145256756978321\\
-30	0.0454429678852096\\
-25	0.139067193338789\\
-20	0.400537929356433\\
-15	1.00857928482707\\
-10	2.0703893253415\\
};
\addlegendentry{Non-aligned best AP}

\addplot [color=blue, dashed, line width=1.0pt, mark=square, mark options={solid, blue}]
  table[row sep=crcr]{%
-40	0.00578734822676539\\
-35	0.0182485601635524\\
-30	0.0571898876072464\\
-25	0.175953268960201\\
-20	0.514552171530819\\
-15	1.34419944793643\\
-10	2.94549018813898\\
};
\addlegendentry{Non-aligned $\mathbf{w}_1 = \mathbf{a}$, $\mathbf{w}_2 = \mathbf{c}$}

\addplot [color=black, dashed, line width=1.0pt, mark=pentagon, mark options={solid, black}]
  table[row sep=crcr]{%
-40	7.71250467758443e-05\\
-35	0.000243883230887079\\
-30	0.00077115346397452\\
-25	0.00243787006000641\\
-20	0.00770204418867143\\
-15	0.0242845238163798\\
-10	0.0760897217241706\\
};
\addlegendentry{Non-aligned $\mathbf{w}_1 = \mathbf{a}$, $\mathbf{w}_2 = \mathbf{b}$}

\end{axis}
\end{tikzpicture}%
    \caption{Achievable rates with different beamforming designs and transmission schemes for the example considered in \eqref{eqn:analyticalExample1} and \eqref{eqn:analyticalExample2}.}
    \label{fig:ApOperation}
\end{figure}

Fig.~\ref{fig:ApOperation} shows the achievable rates in all transmission strategies for the above example. 
For this plot, we consider $M=16,~N=2$, and $\alpha=0.7$. 
The key observations are:
\begin{enumerate}

    \item \glspl{ap} with phase-alignment give the maximum rate. 
    Hence, the phase of \glspl{ap} should be aligned whenever possible to improve data rates. 

    \item The gap between optimal and sub-optimal beamforming significantly varies based on the designs of beamforming and transmission schemes. 
    

\end{enumerate}
	
The above observations give motivations for proposing and optimizing the \gls{pcnc} operation in a general network that contains many \glspl{ap} and \glspl{ue}, which will be considered in the rest of the paper. 
Although not all \glspl{ap} can be phase-aligned, it is possible to synchronize the \glspl{ap} in clusters, and hence, there are phase-aligned clusters in the network. 
A cluster can operate coherently, while different clusters operate non-coherently, thus yielding a \gls{pcnc} scenario. 
Maximizing the sum rate of the network with the \gls{pcnc} operation will strongly require optimizing the beamforming. 
Also, multi-antenna \glspl{ue} require optimizing the combining and data stream allocation to cope with inter-stream interference as discussed in Section~\ref{sec:Introduction}-\ref{sec:relatedwork}.

\section{System Model and Problem Formulation}	\label{sec:SystemModel}

We consider a cell-free massive \gls{mimo} system with $L$ \glspl{ap} and $K$ \glspl{ue}. 
We assume that each \gls{ap} is equipped with $ M $ antennas and each \gls{ue} is equipped with $ N $ antennas. 
All  \glspl{ap} are connected to a \gls{cpu} through high-capacity fronthaul links. 
Let $ \mathbf{G}_{kl} \in \mathbb{C}^{N\times M} $ be the channel matrix between the \gls{ue} $ k \in \K \triangleq \{1,\dots,K\} $ and the \gls{ap} $ l \in \LL \triangleq \{1,\dots,L\} $. 
	
We consider a network with the \gls{pcnc} operation, where the \glspl{ap} are grouped into phase-aligned clusters. 
Each cluster forms a virtual large antenna array to transmit data symbols coherently. 
Different clusters are not phased-aligned, and hence, their transmissions are performed in a non-coherent way. 
Let there be $L_c \leq L $ clusters. 
There are no common \gls{ap} among any pair of clusters. 
We will propose an \gls{ap} clustering algorithm to cluster \glspl{ap} in Section~\ref{sec:PCNC_ApClustering}.

The considered \gls{pcnc} system model is a general system model for the networks with the traditional operations \gls{fc} and FNC.
In particular, $L_c = L$ means that there are no clusters and all \glspl{ap} operate independently, which is the \gls{fnc} operation.
Similarly, $L_c=1$ means that all the \glspl{ap} are in a single phase-aligned cluster, which gives the \gls{fc} operation.

Let $\mathcal{C}_c$ be the set of \glspl{ap} in cluster $c \in \{1,2,\cdots,L_c\}$. 
Let $\q_{kc} \in \mathbb{C}^{d_{kc}\times1}$  be the data symbol vector transmitted by cluster $c$ to \gls{ue} $k$, where $d_{kc}\leq \min (M\Abs{\CCC_c},N)$ is the number of data streams, $\Exp{\q_{kc}}=\mathbf{0}$, $\Exp{\q_{kc}\q_{kc}^{\tH}}=\mathbf{I}_{d_{kc}}$, and $\Exp{\q_{kc}\q_{k'c'}^{\tH}}=\mathbf{0}, \forall k' \neq k, c' \neq c$. 
Note that \gls{ap}s within each cluster send the same data stream to achieve high beamforming gain. The \glspl{ap} from different clusters transmit independent and different data streams. 
Cluster $c$ transmits a data stream toward \gls{ue} $k$ with the collective precoding matrix
\begin{equation}
	\label{eqn:PCNC_Wkc}
	\overline{\mathbf{W}}_{kc} = \Sbrac{\begin{matrix}
			\mathbf{W}_{kl_{c,1}} 
			\\ 
			\mathbf{W}_{kl_{c,2}}
			\\ 
			\vdots 
			\\ 
			\mathbf{W}_{kl_{c,|\CCC_c|}}
	\end{matrix}} \in \C^{M\Abs{\CCC_c} \times d_{kc}},
\end{equation}
where ${\mathbf{W}}_{kl_{c,j}}\in \C^{M \times d_{kc}}$ is the precoding matrix at \gls{ap} $l_{c,j}$ in cluster $\CCC_c$, and $j\in\{1,\dots,\Abs{\CCC_c}\}$. 
The transmitted signal from cluster $c$ is 
\begin{equation}
	\mathbf{x}_c = \sqrt{\rho} \sum_{k=1}^{K} \overline{\mathbf{W}}_{kc} \mathbf{q}_{kc},
\end{equation}
where $\rho$ is the normalized maximum transmit power at each \gls{ap}.
The per-\gls{ap} maximum power constraint is expressed as 
\begin{equation}
	\label{eqn:FC_PC}
	\sum_{k=1}^{K} \trace \Brac{\W_{kl} \W_{kl}^{\tH}} \leq 1, ~ ~ \forall l. 
\end{equation}
	
The received signal $\mathbf{y}_k \in \mathbb{C}^{N\times1}$ at \gls{ue} $k$ is given by 
\begin{align}
	\nonumber
	\mathbf{y}_k & = \sum_{c=1}^{L_c} \overline{\mathbf{G}}_{kc} \mathbf{x}_{c} + \mathbf{n}_k  \\
	\label{eqn:PCNC_yk}
	& = \sqrt{\rho} \sum_{c=1}^{L_c}  \overline{\mathbf{G}}_{kc} \sum_{k'=1}^{K} \overline{\mathbf{W}}_{k'c} \mathbf{q}_{k'c} + \mathbf{n}_k ,
\end{align}	
where $\overline{\G}_{kc} = \Sbrac{\G_{kl_{c,1}},\dots,\G_{kl_{c,\Abs{\CCC_c}}}} \in \mathbb{C}^{N\times M\Abs{\CCC_c}} $ is the collective channel matrix of the \glspl{ap} in cluster $\CCC_c$ to \gls{ue} $k$ and $\mathbf{n}_k$ is an \gls{awgn} vector with \gls{iid} $\mathcal{CN}(0,1)$ entries.
To estimate data symbol $\mathbf{q}_{kc}$, \gls{ue} $k$ applies the combining matrix $\overline{\mathbf{V}}_{kc} \in \mathbb{C}^{N\times d_{kc}}$ as
\begin{align}\label{eqn:PCNC_qkc_hat}
    \nonumber
    &\hat{\mathbf{q}}_{kc} 
    = \overline{\mathbf{V}}_{kc}^\text{H} \mathbf{y}_k 
    \\
    \nonumber
    & = \sqrt{\rho} \overline{\mathbf{V}}_{kc}^\text{H} \overline{\mathbf{G}}_{kc} \overline{\mathbf{W}}_{kc} \mathbf{q}_{kc} +  \sqrt{\rho} \overline{\mathbf{V}}_{kc}^\text{H} \sum_{c'=1,c'\neq c}^{L_c} \overline{\mathbf{G}}_{kc'} \overline{\mathbf{W}}_{kc'} \mathbf{q}_{kc'} 
    \\
    & \qquad + \sqrt{\rho}\overline{\mathbf{V}}_{kc}^\text{H} \sum_{c'=1}^{L_c}  \overline{\mathbf{G}}_{kc'} \!\!\!\!\! \sum_{k'=1,k'\neq k}^{K} \!\!\! \!\! \overline{\mathbf{W}}_{k'c'} \mathbf{q}_{k'c'} 
    + \overline{\mathbf{V}}_{kc}^\text{H} \mathbf{n}_k.
\end{align}
The desired signal for the data stream from cluster $c$ to \gls{ue} $k$ is the first term of \eqref{eqn:PCNC_qkc_hat}, while the inter-cluster and inter-\gls{ue} interference are the second and the third terms of \eqref{eqn:PCNC_qkc_hat}, respectively. 
	
Treating the interference terms as Gaussian noise, the achievable rate $R_{kc}$ for the data stream from cluster $c$ to \gls{ue} $k$ as~\cite{guthy2010efficient,christensen2008weighted}
\begin{equation}\label{eqn:PCNC_Rkc}
	R_{kc}  = \log_2\Abs{\mathbf{I}_{d_{kc}} + \rho \overline{\mathbf{H}}_{kc}^\text{H} \overline{\mathbf{Q}}_{kc}^{-1} \overline{\mathbf{H}}_{kc} },
\end{equation}
where
\begin{align}
    \label{eqn:PCNC_Hkc}
    & \overline{\mathbf{H}}_{kc}  =  \overline{\mathbf{V}}_{kc}^\text{H}\overline{\mathbf{G}}_{kc} \overline{\mathbf{W}}_{kc} , 	\\
    \nonumber
    &	\overline{\mathbf{Q}}_{kc}  = ~ \overline{\mathbf{V}}_{kc}^\text{H} \Bigg( \rho \sum_{c'=1,c'\neq c}^{L_c} \overline{\mathbf{G}}_{kc'} \overline{\mathbf{W}}_{kc'} \overline{\mathbf{W}}_{kc'}^\text{H} \overline{\mathbf{G}}_{kc'}^\text{H} 
    \\
    \label{eqn:PCNC_Qkc}
    & + \rho \sum_{c'=1}^{L_c} \overline{\mathbf{G}}_{kc'} \Big(\sum_{k'=1,k'\neq k}^{K} \!\!\!\!\! \overline{\mathbf{W}}_{k'c'}  \overline{\mathbf{W}}_{k'c'}^\text{H}\Big) \overline{\mathbf{G}}_{kc'}^\text{H} \!+\!     \mathbf{I}_N \Bigg)  \overline{\mathbf{V}}_{kc}.
\end{align}
Since the data streams for each \gls{ue} are independently coded among the clusters, the rate at \gls{ue} $k$ is the sum of the achievable rates of the data streams from all clusters to this \gls{ue}, i.e.,  
\begin{equation}
	R_{k}  = \sum_{c=1}^{L_c} R_{kc}.
\end{equation}
	
We aim to optimize precoding and combining matrices as well as the number of data streams per \gls{ue} to maximize the total sum rate under the per-\gls{ap} maximum transmit power constraints. 
This problem can be formulated as
\begin{align} \label{eqn:PCNC_OptimizationProblem}
	\underset{\Cbrac{\overline{\mathbf{W}}_{kc}}, \{\overline{\V}_{kc}\}, \{d_{kc}\}}{\maximize} 
	& \sum_{k=1}^{K} \sum_{c=1}^{L_c} R_{kc} (\Cbrac{\overline{\mathbf{W}}_{kc}}, \overline{\V}_{kc}, \{d_{kc}\})
	\\
	\nonumber
	\mathrm{subject~to} ~~~~ 
	& \eqref{eqn:FC_PC}.
\end{align}
Problem \eqref{eqn:PCNC_OptimizationProblem}, even for the fixed values of $\{d_{kc}\}$, has a similar mathematical structure as that of the problem~\cite{luo2008dynamic}, which was shown therein to be NP-hard. 
Therefore, it is challenging to obtain a globally optimal solution to Problem \eqref{eqn:PCNC_OptimizationProblem}. 
In this paper, we propose sub-optimal solutions to find beamforming vectors and data stream allocation.
These solutions are discussed in Sections~\ref{sec:ans:DesignCombPrecoder} and \ref{sec:DataStreamAllocation}, respectively. 	

\label{E1C1}
\revision{
In this work, the challenge of phase misalignment is overcome by using the proposed \gls{pcnc} framework.
The \gls{pcnc} framework involves three parts: (i) \gls{ap} clustering to group \glspl{ap} into phase-aligned clusters based on reference distance; (ii) optimizing the precoding and combining to maximize the sum rate for given \gls{ap} clusters; (iii) allocating data streams to further improve the sum rate performance.
Specifically, the \gls{ap} clustering in part (i) makes sure the \glspl{ap} are phase-aligned in their clusters to achieve higher beamforming gains. 
Compared to the ideal \gls{fc} operation (where all the \glspl{ap} are phase-aligned), the \gls{pcnc} framework only requires \glspl{ap} to be phase-aligned in clusters, which is more feasible in practical deployments. 
In return, the \gls{pcnc} operation brings inter-cluster interference to the system because the \glspl{ap} clusters are not mutually phase-aligned. This is the price to pay when overcoming the challenge of phase misalignment using the \gls{pcnc} framework. 
The optimization of precoding/combining and the data stream allocation in parts (ii) and (iii) efficiently manage inter-cluster interference to improve the system performance. 
Problem~\eqref{eqn:PCNC_OptimizationProblem} belongs to part (ii), which aims to manage 
the inter-cluster interference for maximizing the sum rates for given AP clusters obtained by part (i). The sum rates are functions of inter-cluster interference as shown in \eqref{eqn:PCNC_Rkc}.
Thus, the problem of phase misalignment is taken into account by managing the inter-cluster interference. 
}

\section{AP Clustering Algorithm} 
\label{sec:PCNC_ApClustering}
This section discusses how phase-aligned \gls{ap} clusters can be formed for coherent transmission and high data rate. 
Synchronizing the phases of all the \glspl{ap} in the network is practically very challenging as discussed in Section~\ref{sec:Introduction}. 
Despite that, a certain set of \glspl{ap} that are near to one another can be phase-aligned as shown in \cite{vieira2021reciprocity,ganesan2023beamsync}. 
Two \glspl{ap} are guaranteed with an acceptable phase alignment if their distance is within a reference distance $D$, which is assumed to be known. 
	
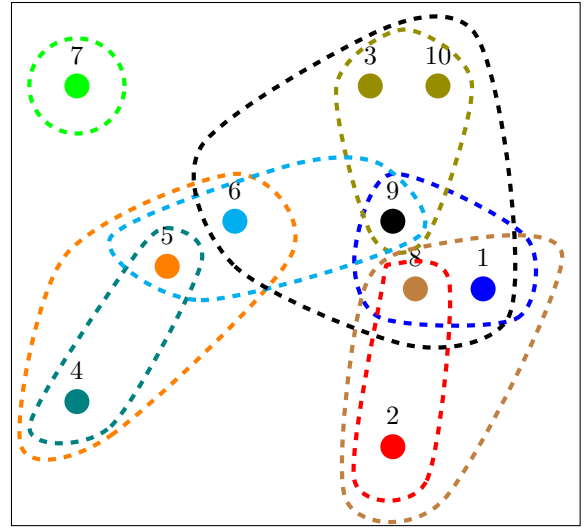
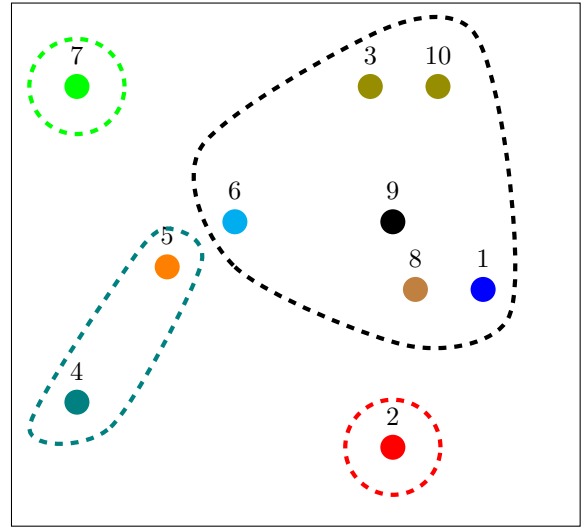
\begin{figure}[!t]
	\centering
	\begin{subfigure}[b]{0.5\textwidth}
		\centering
%
%

%
\begin{tikzpicture}[scale=1.2]
	
%
%

	\draw (-3.1,-3) rectangle (3.2,2.8);

	\node[fill,circle,color=blue ,label=$1$] at (2.125,-0.375){};  
	\node[fill,circle,color=red, label=$2$] at (1.125,-2.125){};  
	\node[fill,circle,color=olive ,label=$3$] at (0.875,1.875){};  
	\node[fill,circle,color=teal, label=$4$] at (-2.375,-1.625){};  
	\node[fill,circle,color=orange, label=$5$] at (-1.375,-0.125){};  
	\node[fill,circle,color=cyan, label=$6$] at (-0.625,0.375){};  
	\node[fill,circle,color=green, label=$7$] at (-2.375,1.875){};  
	\node[fill,circle,color=brown, label=$8$] at (1.375,-0.375){};  
	\node[fill,circle, label=$9$] at (1.125,0.375){};  
	\node[fill,circle,color=olive, label=$10$] at (1.625,1.875){};  
	
	\draw[ultra thick,dashed,name path=left top,color=blue] plot[smooth cycle] coordinates 
	{(1,-0.7) (2.5,-0.7) (2.6,0.2) (1.125,0.9) (0.7,0)};

	\draw[ultra thick,dashed,name path=left top,color=red] plot[smooth cycle] coordinates 
	{(0.7,-2.5) (1.5,-2.5) (1.75,-0.3) (1.125,-0.1) (1,-0.4) };

	\draw[ultra thick,dashed,name path=left top,color=olive] plot[smooth cycle] coordinates 
	{(0.5,1.875) (0.9,0.2) (1.5,0.2) (2,1.9) (1.25,2.5) };

	\draw[ultra thick,dashed,name path=left top,color=teal] plot[smooth cycle] coordinates 
	{(-2.9,-1.9) (-2,-1.9) (-1,-0.1) (-1.375,0.3) (-1.8,-0.1) };

	\draw[ultra thick,dashed,name path=left top,color=orange] plot[smooth cycle] coordinates 
	{(-3,-2.1) (-2.0,-2) (0,0) (-0.6,0.9) (-2.5,0) };	

	\draw[ultra thick,dashed,name path=left top,color=cyan] plot[smooth cycle] coordinates 
	{(-2,0.2) (-1.0,-0.5) (1.4,0.2) (1,1) (0,1) };	
	
	\draw[ultra thick,dashed,name path=left top,color=green] (-2.375,1.875) circle (15pt);
		
	\draw[ultra thick,dashed,name path=left top,color=brown] plot[smooth cycle] coordinates 
	{(0.6,-2.7) (1.7,-2.7) (3,0) (1.5,0.1) (0.8,-0.4) };

	\draw[ultra thick,dashed,name path=left top] plot[smooth cycle] coordinates 
	{(-1,1.2) (-0.625,-0.125) (1.3,-1) (2.4,-0.7) (2.4,1) (2,2.5) (1,2.5)};

\end{tikzpicture}%
		\caption{Phase-aligned \textit{zones}. The zone of an \gls{ap} is marked according to the\\ color of that \gls{ap}.}
        \label{fig:ApZones}
	\end{subfigure}
	\begin{subfigure}[b]{0.5\textwidth}
		\centering
%
%

%
\begin{tikzpicture}[scale=1.2]
	
		%
		%
		
	\draw (-3.1,-3) rectangle (3.2,2.8);	
	
	\node[fill,circle,color=blue ,label=$1$] at (2.125,-0.375){};  
	\node[fill,circle,color=red, label=$2$] at (1.125,-2.125){};  
	\node[fill,circle,color=olive ,label=$3$] at (0.875,1.875){};  
	\node[fill,circle,color=teal, label=$4$] at (-2.375,-1.625){};  
	\node[fill,circle,color=orange, label=$5$] at (-1.375,-0.125){};  
	\node[fill,circle,color=cyan, label=$6$] at (-0.625,0.375){};  
	\node[fill,circle,color=green, label=$7$] at (-2.375,1.875){};  
	\node[fill,circle,color=brown, label=$8$] at (1.375,-0.375){};  
	\node[fill,circle, label=$9$] at (1.125,0.375){};  
	\node[fill,circle,color=olive, label=$10$] at (1.625,1.875){};

	\draw[ultra thick,dashed,name path=left top,color=red] (1.125,-2.125) circle (15pt);

	\draw[ultra thick,dashed,name path=left top,color=teal] plot[smooth cycle] coordinates 
	{(-2.9,-1.9) (-2,-1.9) (-1,-0.1) (-1.375,0.3) (-1.8,-0.1) };
	
	\draw[ultra thick,dashed,name path=left top,color=green] (-2.375,1.875) circle (15pt);

	\draw[ultra thick,dashed,name path=left top] plot[smooth cycle] coordinates 
	{(-1,1.2) (-0.625,-0.125) (1.3,-1) (2.4,-0.7) (2.4,1) (2,2.5) (1,2.5)};

\end{tikzpicture}%
		\caption{Non-overlapping phase-aligned \textit{clusters}.}
        \label{fig:ApClusters}
	\end{subfigure}
	\caption{\gls{ap} Clustering}
	\label{fig:ApClustering}
\end{figure}

Let $D_{ll'}$ be the distance between \glspl{ap} $l$ and $l'$. 
We say that \gls{ap} $l'$ is a neighbor of \gls{ap} $l$ if $D_{ll'} \leq D$. 
Let 
\begin{equation}
    \label{eqn:ApZoneFormation}
	\ZZ_l = \Cbrac{l' | D_{ll'} \leq D, l'\in \Cbrac{1,2,\cdots,L}}
\end{equation}
be the phase-aligned \textit{zone} of \gls{ap} $l$, which includes \gls{ap} $l$ itself and its neighboring \glspl{ap}.
\revision{An example of how zones are formed for each \gls{ap} is provided in Fig.~\ref{fig:ApZones}. 
For each \gls{ap} zones are marked according to the color of the \gls{ap} in the figure. 
Note that, all the \glspl{ap} included in the zone of an \gls{ap} are within the reference distance $D$ and hence, can be phase synchronized.
With zones, we are considering only neighboring \glspl{ap} of an \gls{ap} and hence, the zones can overlap with each other.
}

The \glspl{ap} in a phase-synchronized zone	can perform coherent transmission to a \glspl{ue}. 
Here, the terminology ``\textit{zone}'' instead of ``\textit{cluster}'' is used. 
This is because some \glspl{ap} might share the same phase-synchronized zones and some phase-synchronized zones can overlap with each other, while the clusters are non-overlapping. 
An example of phase-aligned zones and clusters is given in Fig.~\ref{fig:ApClustering}.
 
We propose an \gls{ap} clustering approach to find the non-overlapping phase-aligned \gls{ap} clusters among the phase-aligned zones, which is provided in Algorithm~\ref{algo:PCNC_ApClustering}. 	
Since the phase of \glspl{ap} should be aligned whenever possible for high data rates as discussed in Section~\ref{sec:ApOperation}, the number of \glspl{ap} in a phase-aligned cluster should be as large as possible. 
Therefore, the key idea of Algorithm~\ref{algo:PCNC_ApClustering} is to select a cluster as the phase-aligned zone that has the largest number of \glspl{ap} in each iteration. 

\label{R1C4}
\revision{Let $\ZZ_l^{(n)}$ be the zone of \gls{ap}-$l$ at iteration-$n$. 
For the first iteration, we initialize $\ZZ_l^{(1)} = \ZZ_l$.}
Denote by
\begin{align}
    \label{eqn:updateLmax}
    L_{\max}^{(n)} & = \underset{l}{\max} \Abs{\ZZ_l^{(n)}},
\end{align}
the largest number of \glspl{ap} in all the zones in iteration $n$. 
Let
\begin{align}
    \label{eqn:updateSmax}
    \SSS_{\max}^{(n)} = \Cbrac{\ZZ_l^{(n)} ~|~ \Abs{\ZZ_l^{(n)}}=L_{\max}^{(n)}},
\end{align}
be the set of zones having the same size of $L_{\max}^{(n)}$. 
\revision{For iteration $1$ for Fig.~\ref{fig:ApZones}, $L_{\max}^{(1)} = 6$ and $\SSS_{\max}^{(1)} = \ZZ_9^{(1)} = \Cbrac{9,3,10,6,8,1}$, the zone of \gls{ap}-$9$. 
}
If $|\SSS_{\max}^{(n)}| > 1$, a cluster is selected as the zone in $\SSS_{\max}^{(n)}$ that has the largest desired signal strength to the \glspl{ue}. 
\revision{Hence, the proposed \gls{ap} clustering algorithm focuses on increasing \glspl{ap} in a cluster considering both the reference distance (largest zone) and the \gls{ue} positions (largest signal strength).}

Let $\ZZ_{l^{\star}}^{(n)}$ be the chosen phase-aligned zone in iteration $n$, where $l^{\star}$ is the index of the corresponding \gls{ap}. 
\revision{For the example in Fig.~\ref{fig:ApZones}, there are no other zones with size $6$, and hence, $\CCC_1 = \ZZ_9^{(1)}$ is chosen as cluster in the first iteration.}
Then, the set of phase-aligned zones is updated by removing the \glspl{ap} of the selected zones from the zones that are not selected, i.e., 
\begin{equation}
    \label{eqn:updateZl}
    \ZZ_l^{(n)} = \ZZ_l^{(n-1)} \setminus \Cbrac{l'~|~ l'\in \ZZ_l^{(n-1)}\cap \ZZ_{l^{\star}}^{(n-1)} }. 
\end{equation}
\revision{
Thus, for the example, we have 
$$\ZZ_4^{(2)} = \Cbrac{4,5}, ~ \ZZ_7^{(2)} = \Cbrac{7}, ~ \ZZ_2^{(2)} = \Cbrac{2}.$$
The \gls{ap} clustering algorithm continues with the next iteration with the updated zones.
With the clustering algorithm, we have 
$$\CCC_2 = \Cbrac{4,5}, ~ \CCC_3 = \Cbrac{2}, ~ \CCC_4 = \Cbrac{7}.$$
}

Algorithm~\ref{algo:PCNC_ApClustering} terminates when there are no zones in which the numbers of \glspl{ap} are larger than $1$. 
The output of Algorithm~\ref{algo:PCNC_ApClustering} is a set~$\CCC$ of non-overlapping phase-aligned \gls{ap} clusters with $L_c = \Abs{\CCC}$.

\begin{algorithm}[!t]
    \caption{\gls{ap} clustering in \gls{pcnc} transmission in cell-free massive \gls{mimo} systems}
    \begin{algorithmic}[1]
        \label{algo:PCNC_ApClustering}
        \renewcommand{\algorithmicrequire}{\textbf{Input:}}
        \renewcommand{\algorithmicensure}{\textbf{Initialize:}}
        \REQUIRE Channel matrices $\G_{kl}, ~\forall k,l$. 
        \ENSURE $n=1$, $\ZZ_l^{(1)}=\ZZ_l, ~\forall l$ , $\CCC = \emptyset$
        \STATE Update $L_{\max}^{(1)}$, $\SSS_{\max}^{(1)}$ using \eqref{eqn:updateLmax}, \eqref{eqn:updateSmax}
        \WHILE{$L_{\max}^{(n)} > 1$}
        \IF{$|\SSS_{\max}^{(n)}| = 1$}
        \STATE Update $\CCC = \CCC \cup 
        \SSS_{\max}^{(n)}$. 
        \ELSE
        \FORALL{$\ZZ_{l_i}^{(n)} \in \SSS_{\max}^{(n)}$}
        \STATE  Update $\overline{\G}_{kl_i} = \Sbrac{\G_{kl_{i,1}},\G_{kl_{i,2}},\cdots,\G_{kl_{i,L_{\max}^{(n)}}} }$
        \ENDFOR
        \STATE Update $\CCC = \CCC \cup \ZZ_{l_i^{\star}}^{(n)}$, \\where $l_i^{\star} = \underset{l_i}{\argmax} \sum_{k \in \K} \Norm{{\overline{\G}}_{kl_i}}_F^2$
        \ENDIF
        \STATE Update $n:=n+1$
        \STATE Update $L_{\max}^{(n)}$, $\SSS_{\max}^{(n)}$, $\ZZ_l^{(n)}$ using \eqref{eqn:updateLmax}--\eqref{eqn:updateZl}
        \ENDWHILE
        \FORALL{$l$ such that $\Abs{\ZZ_l^{(n)}} \neq 0 $}
        \STATE Update $\CCC = \CCC \cup \ZZ_{l}^{(n)}$
        \ENDFOR
        \STATE \textbf{Output:} Set $\CCC$ of non-overlapping \gls{ap} clusters 
    \end{algorithmic}
\end{algorithm}

\section{Precoding and Combining Optimization Algorithm for Sum Rate Maximization} \label{sec:ans:DesignCombPrecoder}
For given values of $\Cbrac{d_{kc}}$, the optimization problem of maximizing the sum rate is given by 
\begin{align}\label{eqn:PCNC_OptimizationPblm_dkcGiven}
	\underset{\Cbrac{\overline{\mathbf{W}}_{kc}}, \{\overline{\V}_{kc}\}}{\maximize} ~ ~ 
	& \sum_{k=1}^{K} \sum_{c=1}^{L_c} R_{kc} (\Cbrac{\overline{\mathbf{W}}_{kc}}, \overline{\V}_{kc})	\\
	\nonumber
	\,\,\,\,\,\, \mathrm{subject\,\,to} \,\,\,\,\,\, 
	& \eqref{eqn:FC_PC}.
\end{align}
In this section, we develop a sub-optimal solution to problem \eqref{eqn:PCNC_OptimizationPblm_dkcGiven} that is workable but still provides good performance.
Specifically, we propose a novel algorithm to optimize the design of precoding and combining matrices for a given number of data streams, by tailoring the \gls{wmmse} framework \cite{shi2011iteratively,shi2015secure} to solve \eqref{eqn:PCNC_OptimizationPblm_dkcGiven}.  
The algorithm exploits the relationship between the \gls{mse} and data rate to transform the problem into a more tractable form, which can be solved by a \gls{bcd} method. 
Note that the power constraints \eqref{eqn:FC_PC} are per-\gls{ap}, rather than per-cluster constraints.
Moreover, the cluster sizes differ from cluster to cluster, causing the size of the optimization variable matrices in the objective function to have varying dimensions. 
Therefore, the \gls{wmmse} framework in \cite{shi2011iteratively,shi2015secure} cannot be directly applied to solve \eqref{eqn:PCNC_OptimizationPblm_dkcGiven}.

There are two key properties of the objective function of problem \eqref{eqn:PCNC_OptimizationPblm_dkcGiven} that motivate the use of the \gls{wmmse} framework \cite{shi2011iteratively,shi2015secure}.
These properties are stated in the following propositions. 
\begin{proposition}
\label{proposition:MMSE_Combiner}
For given $\Cbrac{\overline{\mathbf{W}}_{kc}}$, the instantaneous rate $R_{kc} (\overline{\V}_{kc}) = \log_2\Abs{\mathbf{I}_{d_{kc}} + \rho \overline{\mathbf{H}}_{kc}^\mathrm{H} \overline{\mathbf{Q}}_{kc}^{-1} \overline{\mathbf{H}}_{kc} } $ is maximized by the \gls{mmse} combining matrix
\begin{align}
\label{eqn:Combiner_MMSE_exp}
    \nonumber
    & \overline{\V}_{kc}^{\mathrm{MMSE}} \!=\! \sqrt{\rho} \Bigg( \rho \sum_{c'=1,c'\neq c}^{L_c} \overline{\G}_{kc'} \overline{\W}_{kc'} \overline{\W}_{kc'}^\mathrm{H} \overline{\G}_{kc'}^\mathrm{H} 
    \\
    &  + \rho \sum_{c'=1}^{L_c}  \sum_{k'=1}^{K} \overline{\G}_{kc'} \overline{\W}_{k'c'}  \overline{\W}_{k'c'}^\mathrm{H} \overline{\G}_{kc'}^\mathrm{H}   + \mathbf{I}_N \!\! \Bigg) ^{-1} \!\!\!\! \overline{\G}_{kc} \overline{\W}_{kc},	
\end{align}
which results in the achievable rate for \gls{ue} $k$ in a \gls{mimo} broadcast channel given by \eqref{eqn:RateWith_MMSE_exp} on the top of the next page.
\begin{figure*}
    \begin{align}
        \label{eqn:RateWith_MMSE_exp}
        R_{kc} = \log_2 \Bigg| \mathbf{I}_{d_k} \!+\! \rho \W_k^{\mathrm{H}} \G_k^{\mathrm{H}} 
        \Brac{ \rho \!\!\!\! \sum_{c'=1,c'\neq c}^{L_c} \!\!\! \overline{\G}_{kc'} \overline{\W}_{kc'} \overline{\W}_{kc'}^\mathrm{H} \overline{\G}_{kc'}^\mathrm{H} 
        \!+\! \rho \sum_{c'=1}^{L_c}  \sum_{k'=1,k'\neq k}^{K} \!\!\!\!\overline{\G}_{kc'} \overline{\W}_{k'c'}  \overline{\W}_{k'c'}^\mathrm{H} \overline{\G}_{kc'}^\mathrm{H}   + \mathbf{I}_N \!\!}^{-1} \!\!\!\!\!\! \G_k \W_k \Bigg|.
    \end{align}
    \hrulefill
\end{figure*}
\end{proposition}
\vspace{-8mm}
\begin{proof}
    See Appendix. 
\end{proof}
	
\begin{proposition}
	\label{proposition:rateW}
	For given $\overline{\V}_{kc} = \overline{\V}_{kc}^{\mathrm{MMSE}}$, the instantaneous rate $R_{kc}(\{\overline{\W}_{kc}\})$ in \eqref{eqn:RateWith_MMSE_exp} can be written as
	\begin{align}\label{eqn:RateRk_withWk}
		\nonumber 
		& \widetilde{R}_{kc} (\{\overline{\W}_{kc}\}) =  d_{kc} \\
		& ~~~ + \!\!\underset{\overline{\CC}_{kc} \succ \mathbf{0}, \overline{\U}_{kc}}{\max}\!\! \log_2 \Abs{\overline{\CC}_{kc}} - \trace[\overline{\CC}_{kc}\overline{\E}_{kc}(\overline{\U}_{kc},\! \{ \overline{\W}_{kc}\})] ,
	\end{align}
	where $\overline{\CC}_{kc}\in \C^{d_{kc} \times d_{kc}}, \overline{\U}_{kc}\in \C^{d_{kc} \times d_{kc}}$ are additional variables, and
	\begin{align}\label{eqn:PCNC_Ekc}
		\nonumber
		& \overline{\E}_{kc}(\overline{\U}_{kc},\{\overline{\W}_{kc}\}) \\ 
		\nonumber
		& =  ~ (\II_{d_{kc}}-\sqrt{\rho}\overline{\U}_{kc}^{\mathrm{H}}\overline{\G}_{kc}\overline{\W}_{kc})(\II_{d_{kc}}-\sqrt{\rho} \overline{\U}_{kc}^{\mathrm{H}}\overline{\G}_{kc}\overline{\W}_{kc})^{\mathrm{H}} \\
		\nonumber
		& \quad + \overline{\U}_{kc}^{\mathrm{H}} \left( \rho \!\!\!\! \sum_{c'=1,c'\neq c}^{L_c} \!\!\! \overline{\G}_{kc'} \overline{\W}_{kc'} \overline{\W}_{kc'}^\mathrm{H} \overline{\G}_{kc'}^\mathrm{H} \right. \\
		& \qquad \left. + \rho \sum_{c'=1}^{L_c}  \sum_{k'=1,k'\neq k}^{K} \!\!\!\!\!\!\overline{\G}_{kc'} \overline{\W}_{k'c'}  \overline{\W}_{k'c'}^\mathrm{H} \overline{\G}_{kc'}^\mathrm{H}   + \mathbf{I}_N \!\! \right) \!\! \overline{\U}_{kc} 
	\end{align}
	has the same form of the \gls{mse} with combining matrix $\overline{\U}_{kc}$. 
	The optimal values of $\overline{\U}_{kc}$ and $\overline{\CC}_{kc}$ are 
	\begin{align}
		\overline{\U}_{kc}^{\mathrm{opt}} & = \overline{\V}_{kc}^{\mathrm{MMSE}},	\\
		\overline{\CC}_{kc}^{\mathrm{opt}} & = \Brac{\II_{d_{kc}} - \sqrt{\rho}(\overline{\U}_{kc}^{\mathrm{opt}})^{\mathrm{H}}\overline{\G}_{kc}\overline{\W}_{kc}}^{-1}.
	\end{align}
\end{proposition}
\begin{proof}
The proof follows \cite[Lemma 4.1]{shi2015secure}, and hence, is omitted.
\end{proof}
	
The function $\log_2 \Abs{ \overline{\CC}_{kc}} - \trace\Sbrac{\overline{\CC}_{kc}\overline{\E}_{kc}(\overline{\U}_{kc},\{\overline{\W}_{kc}\}) }$ is concave with respect to each variable $\overline{\CC}_{kc}$, $\overline{\U}_{kc}$, and $\{\overline{\W}_{kc}\}$, and hence, is a tractable function. 
Thus, from Propositions~\ref{proposition:MMSE_Combiner} and~\ref{proposition:rateW}, the \gls{mmse} combining matrix $\overline{\V}_{kc}^\text{MMSE}$ is shown not only to maximize the rate of \gls{ue} $k$ from cluster $c$, but also to transform the rate function into a more tractable form \eqref{eqn:RateRk_withWk} with respect to $\Cbrac{\overline{\W}_{kc}}$. 
Using the properties from Propositions~\ref{proposition:MMSE_Combiner} and~\ref{proposition:rateW}, we transform problem \eqref{eqn:PCNC_OptimizationPblm_dkcGiven} into an equivalent problem as follows: 	
\begin{align} \label{eqn:PCNC_EquivalentProblem}
	\nonumber
	\underset{\Cbrac{\overline{\mathbf{W}}_{kc}}, \{\overline{\V}_{kc}\}, \{\overline{\CC}_{kc}\}}{\minimize} ~ ~ 
	& \sum_{k=1}^{K} \sum_{c=1}^{L_c} \left( \trace \Brac{\overline{\CC}_{kc}\E_{kc}(\overline{\V}_{kc},\{\overline{\W}_{kc}\})} \right. \\
	& \qquad \qquad \qquad \left. - \log_2\Abs{\overline{\CC}_{kc}} \right)	\\
	\nonumber
	\mathrm{subject\,\,to} \,\,\,\,\,\,\,\,\,\,\,\,\,\,\, &\eqref{eqn:FC_PC}.
\end{align}

The objective function of problem \eqref{eqn:PCNC_EquivalentProblem} is the weighted sum-\gls{mse} with the weight matrices $\{\overline{\CC}_{kc}\}$. 
It is jointly non-convex over all the block variables $\Brac{\Cbrac{\overline{\W}_{kc}}, \{\overline{\V}_{kc}\}, \{\overline{\CC}_{kc}\}}$, but convex over each block variable $\Cbrac{\overline{\W}_{kc}}, \{\overline{\V}_{kc}\} $ or $ \{\overline{\CC}_{kc}\}$. 
This motivates the use of the \gls{bcd} method to iteratively minimize the weighted sum-\gls{mse} in the objective function of problem \eqref{eqn:PCNC_EquivalentProblem}.
In particular for fixed $(\Cbrac{\overline{\W}_{kc}}, \{\overline{\CC}_{kc}\})$, we update $\{\overline{\V}_{kc}\}$ based on the insights of Proposition~\ref{proposition:MMSE_Combiner} as follows:
\begin{align} \label{eqn:PCNC_updateVkc}
	\overline{\V}_{kc} = \overline{\V}_{kc}^\mathrm{MMSE}	, ~~~ \forall k,c.
\end{align}
For fixed $(\{\overline{\V}_{kc}\},\{\overline{\W}_{kc}\})$, we update $\{ \overline{\CC}_{kc} \}$ using the insights of Proposition~\ref{proposition:rateW} as
\begin{align}\label{eqn:PCNC_updateCkc}
	\overline{\CC}_{kc} = \Brac{ \II_{d_{kc}} - \sqrt{\rho} \overline{\V}_{kc}^{\tH} \overline{\G}_{kc} \overline{\W}_{kc}  }^{-1} ~~~\forall k,c.
\end{align}
For fixed $(\{\overline{\V}_{kc}\},\{\overline{\CC}_{kc}\})$, we update  $\{\overline{\W}_{kc}\}$ by solving the  problem  \eqref{eqn:PCNC_updateWkc} (see the next page). 
\begin{figure*}
	\begin{align} \label{eqn:PCNC_updateWkc}
		\nonumber
		\underset{\Cbrac{\mathbf{W}_{kc}}}{\minimize} ~ ~ 
		& \sum_{k_1=1}^{K} \sum_{c_1=1}^{L_c} \trace \Brac{\overline{\CC}_{k_1c_1} (\II_{d_{k_1c_1}}-\sqrt{\rho}\overline{\V}_{k_1c_1}^{\tH}\overline{\G}_{k_1c_1}\overline{\W}_{k_1c_1})(\II_{d_{k_1c_1}}-\sqrt{\rho}\overline{\V}_{k_1c_1}^{\tH}\overline{\G}_{k_1c_1}\overline{\W}_{k_1c_1})^{\tH}} 	\\ 
		\nonumber
		& + \sum_{k_1=1}^K \sum_{c_1=1}^{L_c} \trace\left(   \rho \overline{\CC}_{k_1c_1} \sum_{c_2=1,c_2\neq c_1}^{L_c} \overline{\V}_{k_1c_1}^{\tH}\overline{\G}_{k_1c_2} \overline{\W}_{k_1c_2} \overline{\W}_{k_1c_2}^\text{H} \overline{\G}_{k_1c_2}^\text{H} \overline{\V}_{k_1c_1} \right.		\\
		& \qquad \qquad \qquad \left. + \rho \overline{\CC}_{k_1c_1} \sum_{c_2=1}^{L_c}  \sum_{k_2=1,k_2\neq k_1}^{K} \overline{\V}_{k_1c_1}^{\tH} \overline{\G}_{k_1c_2} \overline{\W}_{k_2c_2}  \overline{\W}_{k_2c_2}^\text{H} \overline{\G}_{k_1c_2}^\text{H}  \overline{\V}_{k_1c_1} \right)	\\
		\nonumber
		\mathrm{subject\,\,to} \,\,\,	&\eqref{eqn:FC_PC}.
	\end{align}
	\hrulefill
	\vspace{-10pt}
\end{figure*}
We summarize the steps to solve problem \eqref{eqn:PCNC_EquivalentProblem} (or problem \eqref{eqn:PCNC_OptimizationPblm_dkcGiven}) in Algorithm~\ref{algo:PCNC_Solution}.
\begin{algorithm}[!t]
    \caption{Solving problem \eqref{eqn:PCNC_EquivalentProblem}}
	\begin{algorithmic}[1]
		\label{algo:PCNC_Solution}
        \renewcommand{\algorithmicrequire}{\textbf{Input:}}
        \renewcommand{\algorithmicensure}{\textbf{Initialize:}}
		\ENSURE $\{\overline{\W}_{kc}\}$ that satisfies \eqref{eqn:FC_PC} 
		\REPEAT
		\STATE Update $\{\overline{\V}_{kc}\}$ as \eqref{eqn:PCNC_updateVkc}
		\STATE Update $\{\overline{\CC}_{kc}\}$ as \eqref{eqn:PCNC_updateCkc}
		\STATE Update $\{\overline{\W}_{kc}\}$ by solving \eqref{eqn:PCNC_updateWkc} 
		\UNTIL{convergence}
	\end{algorithmic}
\end{algorithm}

Since $\overline{\G}_{kc} \overline{\W}_{kc} = \sum_{l\in \CCC_c} \G_{kl} \W_{kl}, $
the problem \eqref{eqn:PCNC_updateWkc} can be equivalently written as problem \eqref{eqn:PCNC_updateWkc_Equivalent} (see the top of the next page). 
In light of the per-\gls{ap} power constraints in \eqref{eqn:FC_PC}, $\{ \W_{kl} \} $ can be considered as a block variable. 
Therefore, we apply the \gls{bcd} approach with block variables $\{ \W_{kl} \}$, for all $k\in\K, l\in\LL$, to solve problem \eqref{eqn:PCNC_updateWkc_Equivalent}. 
Since problem \eqref{eqn:PCNC_updateWkc_Equivalent} is convex with a single quadratic constraint, we can obtain a closed-form solution by dealing with its dual problem as follows. 
\begin{figure*}
\begin{align} \label{eqn:PCNC_updateWkc_Equivalent}
    \nonumber
    \underset{\Cbrac{\mathbf{W}_{kl}}}{\minimize} ~ ~ 
    \tilde{f}(\Cbrac{\mathbf{W}_{kl}}) & = -\sqrt{\rho} \sum_{k_1=1}^{K} \sum_{c_1=1}^{L_c} \sum_{l_1\in\CCC_{c_1}}  \trace\Brac{ \overline{\CC}_{k_1c_1} \overline{\V}_{k_1c_1}^{\tH} \G_{k_1l_1} \W_{k_1l_1} } \\
    \nonumber
    & \qquad - \sqrt{\rho} \sum_{k_1=1}^{K} \sum_{c_1=1}^{L_c} \sum_{l_1\in\CCC_{c_1}} \trace\Brac{ \W_{k_1l_1}^{\tH} \G_{k_1l_1}^{\tH}  \overline{\V}_{k_1c_1} \overline{\CC}_{k_1c_1}  } \\ 
    \nonumber
    & \qquad + \rho \sum_{k_1=1}^K \sum_{c_1=1}^{L_c} 
    \sum_{c_2=1}^{L_c} 
    \sum_{l_1\in\CCC_{c_2}} \sum_{l_2\in\CCC_{c_2}}
    \trace \Brac{ \W_{k_1l_2}^{\tH} \G_{k_1l_2}^{\tH} \overline{\V}_{k_1c_1} \overline{\CC}_{k_1c_1}  \overline{\V}_{k_1c_1}^{\tH}\G_{k_1l_1} \W_{k_1l_1} } 
    \\
    & \qquad + \rho \sum_{k_1=1}^K \sum_{k_2=1,k_2\neq k_1}^{K} \sum_{c_1=1}^{L_c} \sum_{c_2=1}^{L_c} \sum_{l_1\in\CCC_{c_2}} \sum_{l_2\in\CCC_{c_2}}  \!\!\!
    \trace \Brac{ \W_{k_2l_2}^{\tH} \G_{k_1l_2}^{\tH}   \overline{\V}_{k_1c_1} \overline{\CC}_{k_1c_1}  \overline{\V}_{k_1c_1}^{\tH} \G_{k_1l_1} \W_{k_2l_1}   }
    \\
    \nonumber
    \mathrm{subject\,\,to} \qquad \quad	& \eqref{eqn:FC_PC}.
\end{align}
\hrulefill
\vspace{-12pt}
\end{figure*}	
	
The Lagrangian function associated with problem \eqref{eqn:PCNC_updateWkc_Equivalent} with a Lagrangian multipliers $\Cbrac{\lambda_{l}}$ for the power constraints \eqref{eqn:FC_PC} is defined as 
\begin{align} \label{eqn:PCNC_LagrangianExpression}
	\nonumber
	&\LL(\Cbrac{\mathbf{W}_{kl}},\{\lambda_l\})   \\
	& = \tilde{f}(\Cbrac{\mathbf{W}_{kl}}) \! +\! \sum_{l_1=1}^{L} \lambda_{l_1}  \!\!\Brac{\sum_{k_1=1}^{K} \trace (\mathbf{W}_{k_1l_1}\mathbf{W}_{k_1l_1}^\text{H}) - 1}.
\end{align}
Then, the dual problem of problem \eqref{eqn:PCNC_updateWkc_Equivalent} is
\begin{subequations} \label{eqn:PCNC_updateWkc_dual}
	\begin{align} 
		\underset{\lambda_l}{\maximize} ~ ~ &  \widetilde{\LL}(\lambda_l) \\
		\mathrm{subject\,\,to} \,\,\,\, 	&\lambda_l \geq 0, 
	\end{align}
\end{subequations}
where $\widetilde{\LL}(\lambda_l)$ is the dual function, which is given by 
\begin{equation} \label{eqn:PCNC_updateWkc_dual_function}
	\widetilde{\LL}(\lambda_l) = \underset{\Cbrac{\mathbf{W}_{kl}}}{\minimize} ~ ~ 
	\LL(\Cbrac{\mathbf{W}_{kl}},\lambda_l).
\end{equation}
Problem \eqref{eqn:PCNC_updateWkc_dual_function} is convex and its optimal $\W_{kl}$ can be obtained by taking the partial derivative of $\LL(\Cbrac{\mathbf{W}_{kl}},\{\lambda_l\})$ with respect to $\W_{kl}$ and setting it to zero.	
The optimal $\W_{kl}$ for a given $\lambda_l$ is given by \eqref{eqn:PCNC_OptimalWklstar} (see the next page), where $c^l$ is the index of the cluster where $l$-th \gls{ap} belongs to.
	
\begin{figure*}
	\begin{align} \label{eqn:PCNC_OptimalWklstar}
		\nonumber
		\W_{kl}^\star  &  = \Brac{ \rho \sum_{k_1=1,k_1\neq k}^K \sum_{c_1=1}^{L_c} \G_{k_1l}^{\tH} 	\overline{\V}_{k_1c_1} \overline{\CC}_{k_1c_1}  \overline{\V}_{k_1c_1}^{\tH} \G_{k_1l} + \rho \sum_{c_1=1}^{L_c} \G_{kl}^{\tH} \overline{\V}_{kc_1} \overline{\CC}_{kc_1}  \overline{\V}_{kc_1}^{\tH} \G_{kl} + \lambda_l \II_M}^{-1} \\
		\nonumber
		& \qquad \qquad \times \left( \sqrt{\rho}\G_{kl}^{\tH} \overline{\V}_{kc^l} \overline{\CC}_{kc^l} - \rho \sum_{c_1=1}^{L_c} \sum_{l_1\in \CCC_{c^l}\setminus \{l\} } \G_{kl}^{\tH} \overline{\V}_{kc_1} \overline{\CC}_{kc_1}  \overline{\V}_{kc_1}^{\tH} \G_{kl_1} \W_{kl_1} \right. \\
		& \left. \qquad \qquad  \qquad \qquad - \rho \sum_{k_1=1,k_1\neq k}^K \sum_{c_1=1}^{L_c} \sum_{l_1\in \CCC_{c^l}\setminus \{l\} } \G_{k_1l}^{\tH} \overline{\V}_{k_1c_1} \overline{\CC}_{k_1c_1}  \overline{\V}_{k_1c_1}^{\tH} \G_{k_1l_1} \W_{kl_1} \right)
	\end{align}
	\hrulefill
	\vspace{-10pt}
\end{figure*}
	
The Lagrangian multiplier $\lambda_l \geq 0$ should be chosen to satisfy the complementary slackness conditions (of strong duality) corresponding to the power constraint~\eqref{eqn:FC_PC}, i.e., 
\begin{equation} \label{eqn:FC_slackness}
	\lambda_l \Bigg(\sum_{k=1}^{K} \trace (\mathbf{W}_{kl}^{\star}(\mathbf{W}_{kl}^{\star})^\text{H}) - 1\Bigg) = 0, ~~ \forall l.
\end{equation}
Towards this, we analyze the transmission power constraint using the optimal value $\W_{kl}^{\star}(\lambda_l)$ for given $\lambda_l$ as follows. 

Problem \eqref{eqn:PCNC_updateWkc} is convex \gls{qcqp} and can be solved by a commercial interior-point optimization solver such as CVX~\cite{cvx}. 
However, using CVX to solve \eqref{eqn:PCNC_updateWkc} is computationally demanding, especially when the numbers of \glspl{ap} and \glspl{ue} are large. 
Therefore, in the following, we propose an algorithm to find the solution to \eqref{eqn:PCNC_updateWkc} with closed-form expressions in each iteration. 
The proposed algorithm will require much lower computational resources in a large-scale system than using a commercial convex solver.
\begin{algorithm}[!t]
	\caption{Bisection Search for Solving Problem \eqref{eqn:PCNC_updateWkc_Equivalent}}
	\begin{algorithmic}[1]
		\label{algo:PCNC_BisectionSearchMethod}
		\renewcommand{\algorithmicrequire}{\textbf{Input:}}
		\renewcommand{\algorithmicensure}{\textbf{Initialize:}}
		\REQUIRE $\{\overline{\V}_{kc}\}$, $\{\overline{\CC}_{kc}\}$, $\{\W_{kl}\}$, $\epsilon$. 
		\ENSURE $\{\PSI_{kl}\}$ and $\{\SIGMA_{kl}\}$
		\FOR{$l=1:L$}
		\STATE Update $\Cbrac{\LAMBDA_{kl}}$, $\forall~k$ using \eqref{eqn:PCNC_PSI_SIGMA}
		\STATE Update $\Cbrac{\T_{kl}}$, $\forall~k$ using \eqref{eqn:PCNC_UpdateTkl}
		\STATE Initialize $\lambda_{lb}=0$ and \\$\lambda_{ub} = \sqrt{\sum_{k=1}^{K} \sum_{m=1}^{M} \rho [\T_{kl}]_{mm}}$
		\WHILE{$|\lambda_{ub}-\lambda_{lb}| \leq \epsilon$}
		\STATE Update $\lambda_l = \frac{\lambda_{lb} + \lambda_{ub}}{2}$
		\IF{$\sum_{k'=1}^{K} \trace (\mathbf{W}_{k'l}^{\star}(\lambda_l)(\mathbf{W}_{k'l}^{\star}(\lambda_l))^\text{H}) \geq 1$}
		\STATE set $\lambda_{lb} = \lambda_l$ 
		\ELSE 
		\STATE set $\lambda_{ub} = \lambda_l$
		\ENDIF 
		\ENDWHILE
		\STATE Set $\lambda_l^{\star} = \lambda_l$
		\STATE Update $\{\W_{kl}^{\star}\}$ for given $\lambda_l^{\star}$ as \eqref{eqn:PCNC_OptimalWklstar} $\forall k$.
		\ENDFOR
	\end{algorithmic}
\end{algorithm}

Using the \gls{svd}, \eqref{eqn:PCNC_PSI_SIGMA} and \eqref{eqn:PCNC_LAMBDA} (see the next page) hold. 
\begin{figure*}
	\begin{align}
		\label{eqn:PCNC_PSI_SIGMA}
		\PSI_{kl} \SIGMA_{kl} \PSI_{kl}^{\tH} & = \rho \sum_{k_1=1,k_1\neq k}^K \sum_{c_1=1}^{L_c} \G_{k_1l}^{\tH} \overline{\V}_{k_1c_1} \overline{\CC}_{k_1c_1}  \overline{\V}_{k_1c_1}^{\tH} \G_{k_1l}  + \rho \sum_{c_1=1}^{L_c} \G_{kl}^{\tH} \overline{\V}_{kc_1} \overline{\CC}_{kc_1}  \overline{\V}_{kc_1}^{\tH} \G_{kl} \\
		\label{eqn:PCNC_LAMBDA}
		\LAMBDA_{kl} & =  \sqrt{\rho}\G_{kl}^{\tH} \overline{\V}_{kc^l} \overline{\CC}_{kc^l} - \rho \sum_{c_1=1}^{L_c} \sum_{l_1\in \CCC_{c^l}\setminus \{l\} } \G_{kl}^{\tH} \overline{\V}_{kc_1} \overline{\CC}_{kc_1}  \overline{\V}_{kc_1}^{\tH} \G_{kl_1} \W_{kl_1}  
	\end{align}
	\hrulefill
	\vspace{-10pt}
\end{figure*}
From \eqref{eqn:PCNC_OptimalWklstar} and \eqref{eqn:PCNC_PSI_SIGMA}, the transmission power at \gls{ap} $l$ can be written in terms of $\lambda_l$ as \eqref{eqn:PCNC_PowerConstraintLagragian} (see the next page).
\begin{figure*}
	\begin{align}
		\nonumber
		\sum_{k=1}^{K} \trace (\mathbf{W}_{kl}^{\star}(\mathbf{W}_{kl}^{\star})^\text{H}) 
		&=\sum_{k=1}^{K} \trace ((\mathbf{W}_{kl}^{\star})^\text{H}\mathbf{W}_{kl}^{\star})	\\
		\nonumber
		&= \sum_{k=1}^{K} \trace \Bigg[ \rho \LAMBDA_{kl}^{\tH} \Brac{\rho \PSI_{kl} \SIGMA_{kl} \PSI_{kl}^{\tH} + \lambda_l \II_{M}}^{-1} \Brac{ \rho \PSI_{kl} \SIGMA_{kl} \PSI_{kl}^{\tH} + \lambda_l \II_{M}}^{-1} \LAMBDA_{kl} \Bigg] \\ 
		\label{eqn:PCNC_PowerConstraintLagragian}
		&= \sum_{k=1}^{K} \trace \Sbrac{\rho  \Big( \rho  \SIGMA_{kl}  + \lambda_l \II_{M}\Big)^{-2} \PSI_{kl}^{\tH} \LAMBDA_{kl} \LAMBDA_{kl}^{\tH}  \PSI_{kl}}.
	\end{align}
	\hrulefill
\end{figure*}
Let 
\begin{equation}
	\label{eqn:PCNC_UpdateTkl}
	\T_{kl} = \PSI_{kl}^{\tH} \LAMBDA_{kl} \LAMBDA_{kl}^{\tH} \PSI_{kl} .
\end{equation}
Then, from \eqref{eqn:PCNC_PowerConstraintLagragian} and \eqref{eqn:PCNC_UpdateTkl}, it is true that 
\begin{align} \label{eqn:PCNC_PowerConstraintSimplified}
	\nonumber
	\sum_{k=1}^{K} \trace (\mathbf{W}_{kl}^{\star}(\mathbf{W}_{kl}^{\star})^\text{H}) &  = \sum_{k=1}^{K} \trace \Brac{ \rho  \Big( \rho  \SIGMA_{kl}  + \lambda_l \II_{M}\Big)^{-2} \T_{kl} } \\
	& =  \sum_{k=1}^{K} \sum_{m=1}^{M} \frac{ \rho 
		[\T_{kl}]_{mm}}{(\rho [\SIGMA_{kl}]_{mm} + \lambda_l)^2}.
\end{align}

As seen from \eqref{eqn:PCNC_PowerConstraintSimplified}, the transmit power at \gls{ap} $l$ is a monotonically decreasing function in $\lambda_l$.
Based on this property, we can choose $\lambda_l$ that satisfies the slackness condition in \eqref{eqn:FC_slackness} by a bisection search, i.e., choosing $\lambda_l \geq 0$ to ensure $\sum_{k=1}^{K} \trace (\mathbf{W}_{kl}^{\star}(\mathbf{W}_{kl}^{\star})^\text{H}) - 1 = 0$. 
In the bisection search method, $\lambda_l$ is searched within the range $(\lambda_{lb},\lambda_{ub})$. 
Here, we choose 
\begin{align}
	\label{lambdalb}
	\lambda_{lb} & =0  	\\
	\label{lambdaub}
	\lambda_{ub} & = \sqrt{\sum_{k=1}^{K} \sum_{m=1}^{M} \rho [\T_{kl}]_{mm}},
\end{align}
so that 
\begin{align}
	\sum_{k=1}^{K} \sum_{m=1}^{M} \frac{ \rho [\T_{kl}]_{mm}}{(\rho [\SIGMA_{kl}]_{mm} + \lambda_l)^2} \leq 
	\sum_{k=1}^{K} \sum_{m=1}^{M} \frac{ \rho [\T_{kl}]_{mm}}{\lambda_l^2} \leq 1.
\end{align}

The bisection search algorithm to solve problem  \eqref{eqn:PCNC_updateWkc_Equivalent} is provided in Algorithm~\ref{algo:PCNC_BisectionSearchMethod}.
Algorithm~\ref{algo:PCNC_BisectionSearchMethod} converges to a stationary solution to problem \eqref{eqn:PCNC_updateWkc_Equivalent} (hence, problem \eqref{eqn:PCNC_updateWkc}). 
The proof of the convergence of Algorithm~\ref{algo:PCNC_BisectionSearchMethod} follows a similar proof as in \cite{shi2011iteratively}, and hence, is omitted.

\section{Data Stream Allocation} \label{sec:DataStreamAllocation}
This section will answer the next question of how to optimally allocate the data streams to the \glspl{ue}.
If $ML > KN$, the maximum number of orthogonal data streams that can be transmitted from the effective array constituted by the \glspl{ap} is $KN$, which is ideally achieved in a system with full-rank channels, appropriate combining and precoding designs. 
However, if $KN > ML $, then $KN$ data streams cannot be made orthogonal over the channels, resulting in strong inter-stream interference. 
Therefore, it is important to efficiently allocate data streams to \glspl{ue}, which helps to manage inter-stream interference and improve the sum rate.

In this section, we allocate data streams to the \glspl{ue} in a greedy manner. 
This helps to avoid an exhaustive search over all possible \gls{ue} data stream allocations, which is practically impossible when $L,M,K,N$ are large. 
It has been shown that greedy \gls{ue} scheduling algorithms can provide a performance that is close to the optimum in downlink \gls{mimo} systems with a block diagonalization precoding approach~\cite{tolli2005scheduling,boccardi2007near}.
Therefore, the greedy \gls{ue} scheduling approach is expected to provide good performance.
	
Let $\PP_{kc} $ be the orthonormal basis matrix of $\overline{\G}_{kc}$. 
Then, let
\begin{align}
	\tilde{\G}_{kc} & = \sum_{k'=1,k'\neq k} \PP_{kc}^{\tH} \overline{\G}_{k'c} \\
	\check{\G}_{kc} & = \sum_{c'=1,c'\neq c}^{L_c} \PP_{kc}^{\tH} \overline{\G}_{kc'} 
\end{align}
be the projections of the intra-cluster and inter-cluster interference channels onto the channel $\overline{\G}_{kc}$, respectively. 
Define a matrix $\mathbf{S}\in\mathbb{R}^{K\times L_c}$ whose elements, i.e.,
\begin{align}
	\label{eqn:cinr}
	\Sbrac{\mathbf{S}}_{kc} = \frac{\Norm{\overline{\G}_{kc}}^2}{1 + \Norm{\tilde{\G}_{kc}}^2 + \Norm{\check{\G}_{kc}}^2 } ,
\end{align}
are \gls{cinr}. 
Let $R \Brac{\Cbrac{d_{kc}}} $ be the sum rate obtained from using Algorithm~\ref{algo:PCNC_Solution} to solve problem \eqref{eqn:PCNC_OptimizationPblm_dkcGiven} for a fixed $\Cbrac{d_{kc}}$. 
A greedy data allocation algorithm for sum rate maximization in cell-free massive \gls{mimo} systems is provided in Algorithm~\ref{algo:DataAllocation}. 

The key idea behind the proposed greedy data stream allocation algorithm is to prioritize allocating data streams for the user-cluster pairs with the strongest \glspl{cinr}.
\label{R2C7}\revision{The data stream allocation Algorithm~\ref{algo:DataAllocation} takes the minimum required values of $\DD = \Cbrac{d_{kc}}$ as input. 
To ensure seamless connectivity for all the \glspl{ue}, it is normal to allocate at least one data stream to each \gls{ue}.
Then we initialize the \gls{cinr} matrix according to \eqref{eqn:cinr}.
The algorithm runs for all cluster-user pairs, until data streams are allocated for all the pairs.
}
\revision{In each iteration, we pick the cluster-user pair that has the largest \gls{cinr} in $\mathbf{S}$. 
Denote this pair as $(\hat{k},\hat{c})$.
Then we compute the sum rate as per \eqref{eqn:PCNC_OptimizationPblm_dkcGiven} for different values of data stream, i.e. $d\in \{1,\dots,\min\Cbrac{M,N}\}$, for this cluster-user pair. 
Let 
\begin{align}
\label{eqn:dataStreamAllocationQuantity}
d^* =  \underset{d\in\{1,\dots,\min\Cbrac{M,N}\}}{\argmax} R \Brac{ \mathcal{D}\setminus \Cbrac{d_{\hat{k}\hat{c}} } \bigcup \Cbrac{d_{\hat{k}\hat{c}} = d } },    
\end{align}
be the allocation quantity that maximizes the sum rate. 
We update the data stream allocation quantity for this cluster-user pair to be $d^*$, if $d^*\geq d_{\hat{k},\hat{c}}$, otherwise we allocate the minimum required quantity.  
Then, we remove this cluster-user pair from the matrix $\mathbf{S}$, and the allocation proceeds with the next strongest pair in matrix $\mathbf{S}$. 
}

\begin{algorithm}[!t]
	\caption{Greedy data stream allocation for sum rate maximization.}
	\begin{algorithmic}[1]
		\label{algo:DataAllocation}
		\renewcommand{\algorithmicrequire}{\textbf{Input:}}
		\renewcommand{\algorithmicensure}{\textbf{Initialize:}}
		\REQUIRE $\mathcal{D} = \Cbrac{d_{kc}} $, where $\{d_{kc}\}$ are minimal values required from the system.
		\ENSURE $ \mathbf{S} $ according to \eqref{eqn:cinr}
		\FOR{$t=1:KL_c$}
		\STATE Update $\hat{k},\hat{c} = \underset{k,c}{\argmax} ~ \Sbrac{\mathbf{S}}_{kc}$
		\STATE Compute $d^*$ as per \eqref{eqn:dataStreamAllocationQuantity}
        \IF{$ d^* \geq d_{\hat{k}\hat{c}}$ }
        \STATE $d_{\hat{k}\hat{c}} = d^*$
        \ENDIF
        \STATE Update $\Sbrac{\mathbf{S}}_{\hat{k}\hat{c}} = 0$
		\ENDFOR
	\end{algorithmic}
\end{algorithm}

\section{Complexity Analysis}
\label{sec:ComplexityAnalysis}
\revision{In this section, we provide a complexity analysis of the proposed algorithms. }

\subsection{AP Clustering Algorithm}
\revision{Algorithm \ref{algo:PCNC_ApClustering} for \gls{ap} clustering has low complexity because it involves only the computation of distances and Frobenius norms of channel matrices. 
The complexity for the computation of distances between \glspl{ap} has a complexity of $\OO(L^2)$.
The system has $L$ \glspl{ap} and $K$ \glspl{ue} and thus, the Frobenius norm involves $\OO(KLMN)$ computations.
Hence, the total complexity for Algorithm~\ref{algo:PCNC_ApClustering} is $\OO(KLMN + L^2)$.  
}

\subsection{Precoding and Combining Optimization Algorithm}
\revision{
Algorithm~\ref{algo:PCNC_Solution} for the
precoding and combining optimization involves the computation of the matrices $\overline{\V}_{kc}$, $\overline{\CC}_{kc}$, and $\overline{\W}_{kc}$ for each cluster-user pair in each iteration. 
The computation of the \gls{mmse} matrix $\overline{\V}_{kc}$ in~\eqref{eqn:PCNC_updateVkc} consists of an $N\times N$ matrix inversion and some matrix multiplications. 
The complexity of computing $\overline{\mathbf{V}}_{kc}$ involves computations of $\OO(KL_c(N^3 + N^2ML + N^2d + NMLd ))$, where $d$ is the maximum number of data streams allocated.
The complexity of computing $\overline{\mathbf{C}}_{kc}$ in~\eqref{eqn:PCNC_updateCkc} involves computations of $\OO(KL_c(d^3 + N^2d + NMLd ))$. 
To compute $\overline{\W}_{kc}$, we use Algorithm~\ref{algo:PCNC_BisectionSearchMethod} with closed-form expressions. 
The computation of $\LAMBDA_{kl}$ requires $\OO(LMNd + LMd^2 + LM^2d)$ operations. 
Thus, for all \glspl{ue} and \glspl{ap}, the total complexity is of$\OO(KL^2MNd + KL^2Md^2 + KL^2M^2d))$.
Similarly, the complexity of computing a$\Cbrac{\PSI_{kl}}$ is $\OO(KL(M^3 + KL_cMNd + KL_cMd^2 + KL_cM^2d))$.
The bisection search completes in a few iterations and thus Algorithm~\ref{algo:PCNC_BisectionSearchMethod} requires $L$ times the computation of $\Cbrac{\LAMBDA_{kl}}$ and $\Cbrac{\PSI_{kl}}$.
Thus, the total complexity of an iteration of Algorithm~\ref{algo:PCNC_Solution} is $\OO(KL_cN^3 + KL_cN^2ML + KL_cN^2d + KL_cNMLd  + KL_cd^3 + KL_c N^2d + KL_c NMLd s + KL^3MNd + KL^3Md^2 + KL^3M^2d + KL^2M^3 + K^2L^2L_cMNd + K^2L^2L_cMd^2 + K^2L^2L_cM^2d)$.
}

\subsection{Data Stream Allocation Algorithm}
\revision{
Algorithm \ref{algo:DataAllocation} for greedy data stream allocation involves the computation of the \gls{cinr} matrix, which has the complexity of $\OO(KLcMNL)$, and running Algorithm~\ref{algo:PCNC_Solution} whose complexity is discussed in the previous subsection.  
}


\section{Numerical Results}	\label{sec:Results}
We consider a cell-free massive \gls{mimo} system, where the \glspl{ap} and \glspl{ue} are randomly distributed in a $0.5~\text{km} \times 0.5~\text{km}$ square area, whose edges are wrapped around to avoid boundary effects. 
The distances between adjacent \glspl{ap} are at least $50$~m~\cite{bjornson2020making}. 
We set the bandwidth to $B=50$~MHz and the noise figure to $F = 9$~dB. 
Thus, the noise power $\sigma_n^2=k_B T_0 B F$, where $k_B=1.381\times 10^{-23}$~Joules/${}^o$K is  Boltzmann's constant, while $T_0=290^o$K is the noise temperature. 
Let $P_d = 1$~W be the maximum transmit power of the \glspl{ap}. 
The normalized maximum transmit power ${\rho}$ is calculated by dividing $P_d$ by the noise power. 
	
The channels are modeled as uncorrelated Rayleigh fading. 
More specifically, $\mathbf{G}_{kl} = \sqrt{\beta_{kl}} \widetilde{\G}_{kl}$, where $\beta_{kl}$ is the large-scale fading coefficient, and $\widetilde{\G}_{kl}$ are \gls{iid} $\CN (0,1)$ variables representing the small-scale fading coefficients.
We model the large-scale fading coefficients $\beta_{kl}$ as~\cite{bjornson2020making}
\begin{align}\label{eqn:fading:large}
	\beta_{kl} = 10^{\frac{\text{PL}(d_{kl})}{10}} 10^{\frac{F_{kl}}{10}},
\end{align}
where
\begin{align}\label{eqn:PL:model}
	\text{PL}(d_{kl})~\text{(dB)} = -30.5-36.7\log_{10}\left(\frac{d_{kl}}{1\,\text{m}}\right),
\end{align}
represents the path loss, and $F_{kl}\in\mathcal{N}(0,4^2)$ (dB) represents the shadowing effect.
The correlation among the shadowing terms from the \gls{ap}~$l, \forall l\in\LL$ to different \glspl{ue} $k \in\K$ is expressed as:
\begin{align}\label{eqn:corr:shadowing}
	\mathbb{E}\{F_{kl}F_{k'l'}\} \triangleq
	\begin{cases}
		4^2 2^{-\delta_{kk'}/9\,\text{m}},& \text{if $l'=l$}\\
		0, & \mbox{otherwise},
	\end{cases}, \forall l'\in\LL,
\end{align}
where $\delta_{kk'}$ is the physical distance between \glspl{ue} $k$ and $k'$.
\label{R1C5}\revision{
In this work, we assume that the channel coefficients are perfectly known at the \glspl{ap} and \gls{cpu}.
In practice, the channels are estimated using pilots.  
For example, a system using time-division duplexing operation has two phases: (i) an uplink transmission phase for channel estimation and (ii) a downlink transmission phase. 
In the uplink transmission phase, all the \glspl{ue} transmit their pilots to all the \glspl{ap}, and then the \glspl{ap} use the pilot signal to estimate the channels. 
When the coherence interval is sufficiently long, the channel estimation is nearly perfect, which is the case considered in this work. 
Here, the \gls{pcnc} operation combining the coherent and non-coherent transmissions is performed in the downlink transmission phase, and hence, does not have any impact on the channel estimation. 
}

\label{R2C8}
\revision{For simulations, we consider $D=200$~m such that there is approximately a $10\%$ drop of the maximum sum rate compared with that of the ideal \gls{fc} scheme. Note that this is an example value of $D$, while the exact value of $D$ depends on used phase-synchronization technologies. }
	
\begin{figure}[!t]
	\centering
	\input{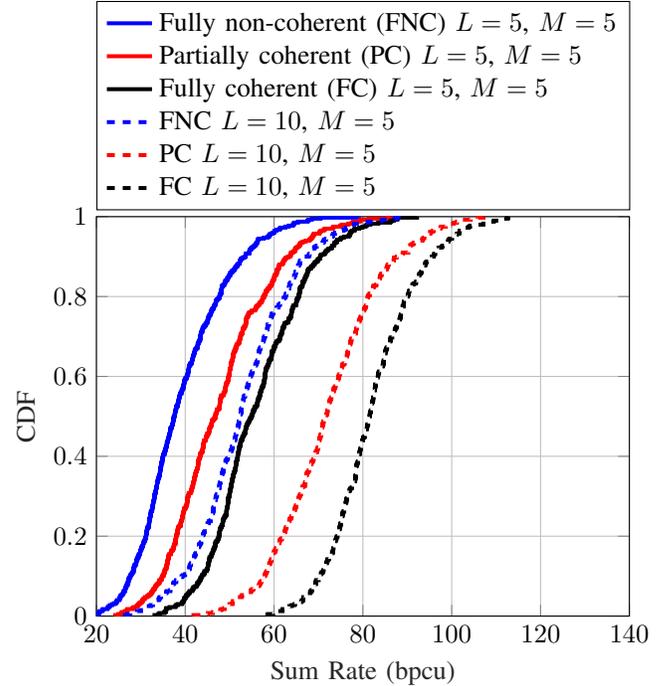} 
	\caption{Comparison of \gls{fnc}, \gls{pcnc}, and \gls{fc} under different network settings. Parameters for the plot: $K=5$, $N=2$, $D=200$~m, and $d_{kc}=2 ~\forall k,c$.}
	\label{fig:sumRateComparison}
\end{figure}

Fig.~\ref{fig:sumRateComparison} shows a sum rate comparison between \gls{fc}, \gls{pcnc}, and \gls{fnc} under different network settings. 
Our proposed \gls{pcnc} scheme can achieve much better performance than the \gls{fnc} scheme. 
Moreover, as the number of \glspl{ap} in the network increases, it can achieve performance close to the \gls{fc} scheme. 
This is reasonable because the \gls{pcnc} scheme has higher beamforming gains obtained from the phase-aligned clusters, compared to the \gls{fc} operation without any phase alignment among the \glspl{ap} in the networks. 

\begin{figure}[!t]
	\centering
	\input{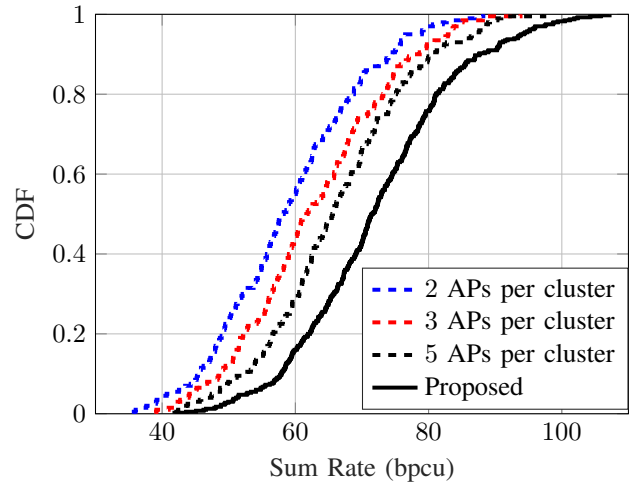} 
	\caption{\revision{Performance of the proposed \gls{ap} clustering algorithm. Parameters for the plot: $L=10$, $K=5$, $M=5$, $N=2$, $D=200$~m, and $d_{kc}=2 ~ \forall k,c$.}}
	\label{fig:clusteringAlgoEfficiency}
\end{figure}

\label{R2C5}
\revision{To show the superiority of our proposed \gls{ap} clustering algorithm, we compare our algorithm with an even distribution of \glspl{ap} among the clusters; see  Fig.~\ref{fig:clusteringAlgoEfficiency}.
With even distribution, we cluster the \glspl{ap} based on the shortest distance between each other. Moreover, the \glspl{ap} are within the reference distance $D$ of the phase-alignment. 
Those \glspl{ap} which are beyond the reference distance operate independently and are not assigned any cluster.
From the figure, it can be seen that the proposed \gls{ap} clustering algorithm outperforms the baseline schemes with an even clustering of \glspl{ap}. 
This is due to the fact that the proposed \gls{ap} clustering algorithm maximizes the channel power in a cluster along with assigning a maximum number of \glspl{ap} per cluster. 
}

\begin{figure}[!t]
	\centering
	\input{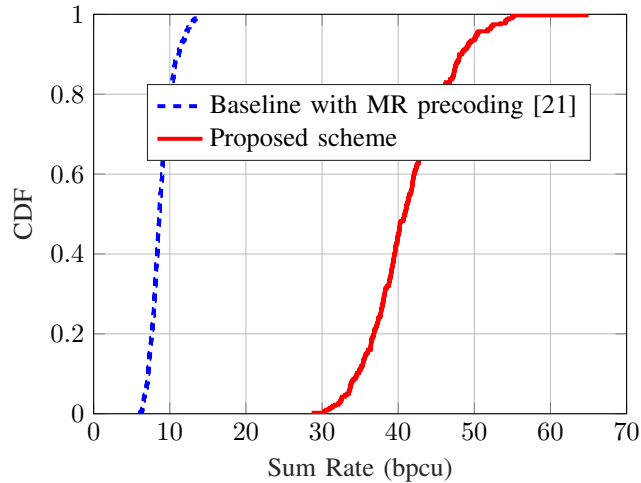} 
	\caption{\revision{Performance of \gls{pcnc} scheme compared to scheme in~\cite{antonioli2023mixed}. Parameters for the plot: $L=10$, $M=5$, $K=5$, and $N=1$.}}
	\label{fig:plotComparisonWithMixed}
\end{figure}

\label{R1C2}
\revision{
In Fig.~\ref{fig:plotComparisonWithMixed}, we compare the proposed scheme with a baseline using \gls{mr} precoding scheme as in~\cite{antonioli2023mixed} for a \gls{pcnc} system. 
For fair comparison with~\cite{antonioli2023mixed}, in this figure, we consider single antenna \glspl{ue} and allocate only a single data stream for every cluster-user pair. 
Here, the baseline uses the same \gls{ap} clustering as that of the proposed \gls{pcnc} scheme.
Note that~\cite{antonioli2023mixed} assumes that the \gls{ap} clusters are known, and does not take into account the practical problem of phase misalignment. Also, \cite{antonioli2023mixed} considers  \gls{mr} precoding and does not consider the aspect of optimizing the beamforming to achieve maximum data rates for the \gls{pcnc} operation.
As discussed in Sec.~\ref{sec:ApOperation}.\ref{sec:PhaseMisAlignedScenario}.\ref{sec:subOptimalBeamforming}, a fixed beamforming technique without considering the channel of the whole network performs poorly as can be seen in the figure. 
Figure~\ref{fig:plotComparisonWithMixed} shows that the \gls{pcnc} scheme obtains a significantly higher sum rate than the baseline, which confirms the importance of optimizing precoding and combining in a \gls{pcnc} system.
}

\begin{figure}[!t]
	\centering
%
%
\begin{tikzpicture}

\begin{axis}[%
width=0.4\textwidth,
height=0.3\textwidth,
at={(0.947in,0.582in)},
scale only axis,
xmin=0,
xmax=450,
xlabel style={font=\color{white!15!black}},
xlabel={Reference Distance $D$ (m)},
ymin=30,
ymax=110,
ylabel style={font=\color{white!15!black}},
ylabel={Mean Sum Rate (bpcu)},
axis background/.style={fill=white},
xmajorgrids,
ymajorgrids,
legend style={at={(0.2,0.02)}, anchor=south west, legend cell align=left, align=left, draw=white!15!black, legend columns=3}
]
\addplot [color=black, line width=1.5pt]
table[row sep=crcr]{%
	10	64.0214440316585\\
	50	69.702935802092\\
	100	79.4636868025725\\
	150	90.8100037674485\\
	200	94.2896462252917\\
	250	97.2084175938696\\
	300	97.9856276784136\\
	350	100.470147355189\\
	400	100.192337783287\\
	450	99.9844595413397\\
	500	100.067585678245\\
};
\addlegendentry{$L=20$}

\addplot [color=blue, line width=1.5pt]
table[row sep=crcr]{%
	10	52.513295277994\\
	50	54.1002237334148\\
	100	60.3801560129256\\
	150	66.4867168271942\\
	200	71.7060216392172\\
	250	75.6205647934834\\
	300	78.3872463956013\\
	350	79.8329635142321\\
	400	81.4104154316921\\
	450	81.4424039673196\\
	500	81.3596196646584\\
};
\addlegendentry{$L=10$}

\addplot [color=red, line width=1.5pt]
table[row sep=crcr]{%
	10	39.0834459632426\\
	50	39.7720374626186\\
	100	40.9114479054366\\
	150	44.9112856474276\\
	200	46.4774516930839\\
	250	50.2184459727402\\
	300	52.0758545217496\\
	350	55.0909826363283\\
	400	55.2747116862167\\
	450	56.7211196788814\\
	500	55.5150188674473\\
};
\addlegendentry{$L=5$}

\addplot [color=black, dashed, forget plot]
  table[row sep=crcr]{%
0	100.527176814894\\
500	100.527176814894\\
};
\addplot [color=blue, dashed, forget plot]
  table[row sep=crcr]{%
0	81.4293808230746\\
500	81.4293808230746\\
};
\addplot [color=red, dashed, forget plot]
  table[row sep=crcr]{%
0	56.6258327885455\\
500	56.6258327885455\\
};

\end{axis}

\end{tikzpicture}%
	\caption{Sum rate of \gls{pcnc} scheme with respect to the reference distance $D$. The dashed lines represent the sum rates of the \gls{fc} operation. \revision{
 Parameters for the plot: $M=5$, $K=5$, $N=2$, and $d_{kc} = 2~\forall k,c$.}}
	\label{fig:PCNC_WithDistance}
\end{figure}
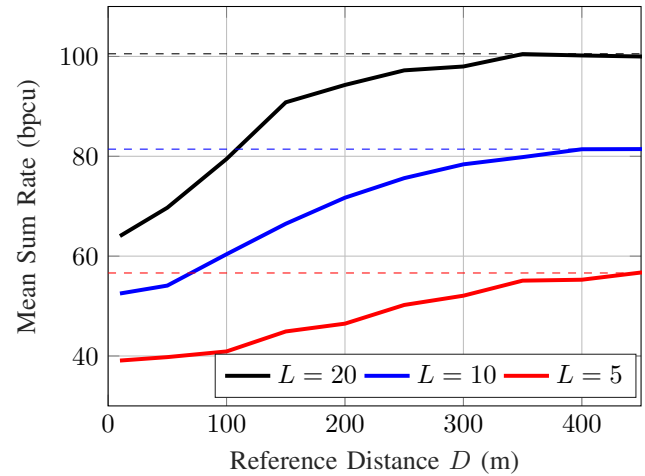
	
Fig.~\ref{fig:PCNC_WithDistance} shows the performance of the \gls{pcnc} system with \gls{ap} clustering under difference values of the reference distance $D$. 
Note that the more $D$ increases, the more \gls{pcnc} system becomes a \gls{fc} system. 
\revision{The larger the reference distance is, the less inter-cluster interference or the price of overcoming phase misalignment (by \gls{ap} clustering) is.}
This is confirmed in Fig.~\ref{fig:PCNC_WithDistance}, where the sum rate saturates after a certain value of the reference distance $D$. 
For a $500\times500~\text{m}^2$ system, it can be seen that forming phase-aligned clusters with approximately up to the reference distance of only $300$~m is needed to achieve a sum rate that is significantly close to that of the system with the \gls{fc} operation. 
Thus, from a system design perspective, it is possible to achieve a near-maximum sum rate of the network without the phase alignment of all \glspl{ap} as in the \gls{fc} operation. 
A \gls{pcnc} system with an appropriate reference distance as well as optimized precoding/combing and data stream allocation would suffice to achieve approximately the sum rate of the \gls{fc} system. 

\begin{figure}[!t]
	\centering
	\input{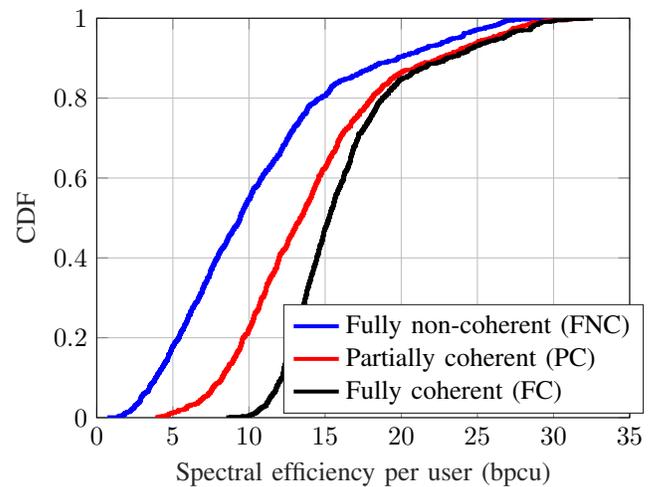} 
	\caption{\revision{Spectral efficiency per user for different transmission schemes. Parameters for the plot: $L=10$, $M=5$, $K=5$, $D=200$~m, and $d_{kc}=2 ~ \forall k,c$. }}
	\label{fig:spectralEfficiencyPerUser}
\end{figure}

\label{R1C3}
\revision{
The spectral efficiency per user for different schemes in the paper is shown in Fig.~\ref{fig:spectralEfficiencyPerUser}.
The \gls{pcnc} scheme performs significantly better than the \gls{fnc} scheme and is close to the \gls{fc} system. 
The \gls{pcnc} system achieves better beamforming gain from the phase-aligned clusters and hence higher rates than the \gls{fnc} system.
Note that the optimization objective considered in this paper is the sum rate of the network. 
Therefore, the difference in the rates of \glspl{ue} over the network can be large. 
}

\begin{figure}[!t]
	\centering
%
%
\begin{tikzpicture}

\glsreset{fc}
\glsreset{fnc}
\glsreset{pcnc}

\begin{axis}[%
width=0.4\textwidth,
height=0.3\textwidth,
at={(0.758in,0.481in)},
scale only axis,
xmin=1,
xmax=7,
xlabel style={font=\color{white!15!black}},
xlabel={Number of \gls{ue} Antennas, $N$},
ymin=20,
ymax=200,
ylabel style={font=\color{white!15!black}},
ylabel={Mean Sum Rate (bpcu)},
axis background/.style={fill=white},
xmajorgrids,
ymajorgrids,
legend style={at={(0.315,0.01)}, anchor=south west, legend cell align=left, align=left, draw=white!15!black}
]
\addplot [color=blue, line width=1.5pt, mark=diamond, mark options={solid, blue}]
  table[row sep=crcr]{%
1	44.5997979981036\\
2	81.7246278154675\\
3	111.482719130177\\
4	137.704013046497\\
5	155.390382429927\\
6	163.168832610404\\
7	170.075733346402\\
8	175.855737235572\\
9	179.1086107105\\
10	183.46190213523\\
};
\addlegendentry{\Gls{fc}}

\addplot [color=red, line width=1.5pt, mark=o, mark options={solid, red}]
  table[row sep=crcr]{%
1	41.5579489492365\\
2	71.2832792845739\\
3	93.5664459324331\\
4	113.071575865167\\
5	124.974478127247\\
6	132.588476135745\\
7	138.644990115446\\
8	143.885259249604\\
9	148.002556456283\\
10	151.106131354696\\
};
\addlegendentry{\Gls{pcnc}}

\addplot [color=black, line width=1.5pt, mark=square, mark options={solid, black}]
  table[row sep=crcr]{%
1	33.7047072156009\\
2	52.5194704667796\\
3	68.7494093024092\\
4	85.605234018086\\
5	98.5149274638232\\
6	108.079166642658\\
7	116.285073278026\\
8	126.727815915101\\
9	132.967532394718\\
10	139.557692295713\\
};
\addlegendentry{\Gls{fnc}}

\end{axis}

\end{tikzpicture}%
	\caption{\revision{Comparison of \gls{fnc}, \gls{pcnc} and \gls{fc} scheme with multiple antennas at the \glspl{ue}. Parameters for the plot: $L=10$, $M=5$, $K=5$, $D=200$~m, and $d_{kc}=\min(M,N) ~ \forall k,c$. }}
	\label{fig:meanSumRateComparisonWithN}
\end{figure}
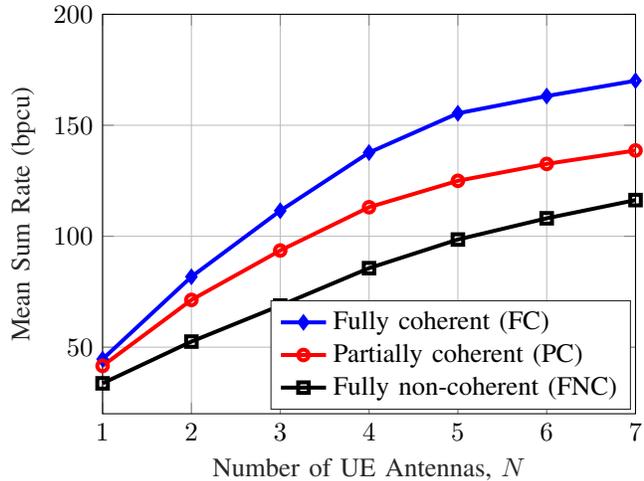

The sum rate performance of cell-free massive \gls{mimo} with multiple-antenna users is provided in Fig.~\ref{fig:meanSumRateComparisonWithN}. 
The sum rate increases significantly with the number of \gls{ue} antennas as the number of data streams allocated to each user increases, until $d_{kc}=\min(M,N) ~ \forall k,c$, beyond which the rate of increase is less. 
This is reasonable because higher beamforming gain is achieved with a larger number of data streams allocated to each user. 
However, a very large number of data streams allocated to each user leads to high inter-stream interference.

\begin{figure}[!t]
	\centering
	\input{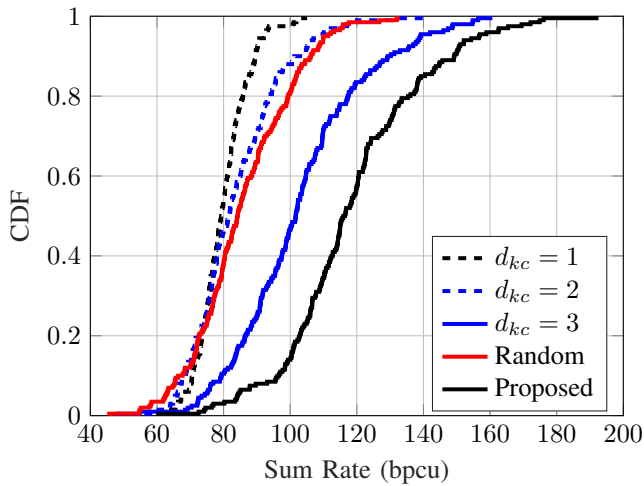} 
	\caption{\revision{Performance of data stream allocation algorithm compared with other methods in a \gls{pcnc} scenario. Parameters for the plot: $L=10$, $M=3$, $K=10$, $N=4$, and $D=200$~m.} }
	\label{fig:dsAllocationEfficiency}
\end{figure}

\revision{
To show the effectiveness of our proposed data stream algorithm in \gls{pcnc} scenario, we compare our algorithm with two additional heuristic baselines: (i) even data stream allocation where every user-cluster pair is allocated the same number of data streams, i.e., $d_{kc} = d,~\forall k,c$; (ii) random data stream allocation where the numbers of data streams allocated for each user-cluster pair are selected randomly, i.e., $d_{kc} = \mathcal{U}(1,\min(M,N))$.
The performance is plotted in Fig.~\ref{fig:dsAllocationEfficiency}. 
From the figure, it can be seen that our proposed algorithm performs significantly better than the other approaches.
}

\begin{figure}[!t]
	\centering
	\input{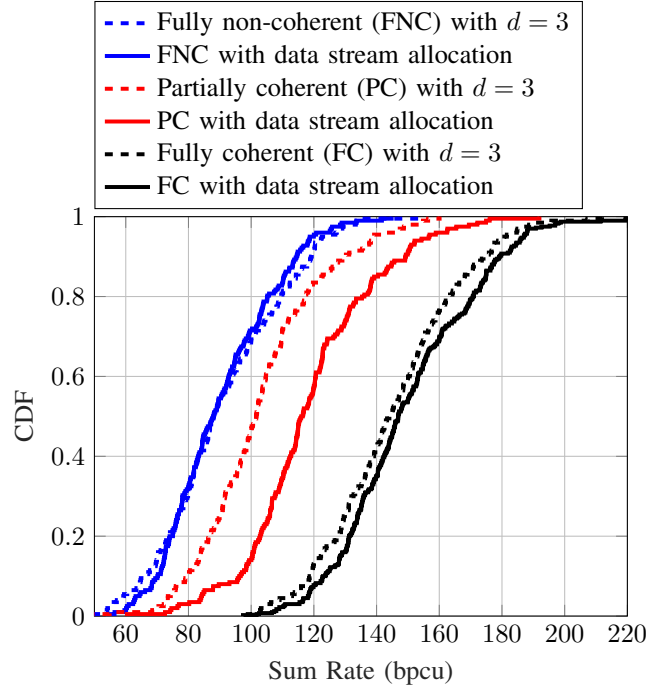} 
	\caption{\revision{Performance comparison of \gls{fnc}, \gls{pcnc}, and \gls{fc} schemes with data allocation. Parameters for the plot: $L=10$, $M=3$, $K=10$, $N=4$, and $D=200$~m.} The dashed curve represents the performance with data allocation and the solid curve represents the performance with fixed data allocation.}
	\label{fig:sumRateWithDataAllocation}
\end{figure}
	
The performance of greedy data stream allocation for all schemes is presented in Fig.~\ref{fig:sumRateWithDataAllocation}. 
From the figure, it can be seen that the improvement in the performance of \gls{pcnc} with data stream allocation as compared to fixed data allocation is much higher than for \gls{fnc} and \gls{fc}.
This highlights the importance of data stream allocation in the \gls{pcnc} operation.

\section{Conclusion} 
\label{sec:Conclusion}
In this paper, we have proposed a novel framework for the \gls{pcnc} operation of cell-free massive \gls{mimo} networks with multi-antenna \glspl{ue}, when there is phase misalignment between the \glspl{ap}. 
In the proposed framework, the subsets of \glspl{ap} form phase-aligned clusters that work together in a non-coherent manner. 
We have proposed an algorithm for clustering \glspl{ap} based on the reference distance of phase alignment. 
We have also developed algorithms to optimize the precoding and combining matrices to maximize the network sum rate, for a given allocation of data streams. 
We also proposed a greedy data-stream allocation algorithm, which improves the sum rate of the \gls{pcnc} operation. 
Numerical results showed that the \gls{pcnc} scheme can obtain significantly higher rates than that of \gls{fnc} operation. 
The proposed \gls{pcnc} can offer a sum rate that is significantly close to that of the \gls{fc} operation, without the need for network-wide phase alignment of the \glspl{ap}. 
This highlights that the \gls{pcnc} operation is a promising solution to enable the practical deployment of cell-free massive \gls{mimo} technology in future communication systems.

\section*{Acknowledgement}
The computations/data handling were enabled by resources provided by the National Academic Infrastructure for Supercomputing in Sweden (NAISS) at Link\"opings Universitet partially funded by the Swedish Research Council through grant agreement no. 2022-06725.
	
\appendix
\section{Proof of Proposition~\ref{proposition:MMSE_Combiner} }	\label{sec:proof_prop1}
Let \eqref{eqn:PCNC_Qkc} be written as 
\begin{align}
	\overline{\mathbf{Q}}_{kc}  = ~ \overline{\mathbf{V}}_{kc}^\text{H}  \A_{kc}  \overline{\mathbf{V}}_{kc},
\end{align}
where $\A_{kc}$ is the second-order interference term given by
\begin{align}
    \label{eqn:PCNC_Qkc_IntfTerm}
    \nonumber
    \A_{kc} & = \mathbf{I}_N  + \rho \sum_{c'=1,c'\neq c}^{L_c} \overline{\mathbf{G}}_{kc'} \overline{\mathbf{W}}_{kc'} \overline{\mathbf{W}}_{kc'}^\mathrm{H} \overline{\mathbf{G}}_{kc'}^\mathrm{H}
    \\
    & \quad + \rho \sum_{c'=1}^{L_c} \overline{\mathbf{G}}_{kc'} \Brac{\sum_{k'=1,k'\neq k}^{K} \!\!\!\!\! \overline{\mathbf{W}}_{k'c'}  \overline{\mathbf{W}}_{k'c'}^\text{H}} \overline{\mathbf{G}}_{kc'}^\mathrm{H} .
\end{align}
It can be shown that the optimal combining matrix that maximizes the instantaneous rate $R_{kc}$ is $\widetilde{\V}_{kc} = \sqrt{\rho} \A_{kc}^{-1}\overline{\G}_{kc} \overline{\W}_{kc}$~\cite[Appendix C.3.2]{bjornson2017massive}. 
Each column of $\widetilde{\V}_{kc}$ is the optimal combining vector for one data stream of \gls{ue} $k$ from cluster $c$. 
The structure of $\widetilde{\V}_{kc}$ can be intuitively explained as follows. 
To detect the data symbol $\q_{kc}$ from \eqref{eqn:PCNC_yk}, we first whiten the inter-\gls{ue} interference plus noise in the received signal $\y_k$ and obtain $\A_{kc}^{-1/2}\y_k$. 
After whitening, the desired signal is highest in the spatial direction $\A_{kc}^{-1/2}\overline{\G}_{kc} \overline{\W}_{kc}$, while the interference plus noise is lowest in this direction. 
Following the \gls{mrc} approach, the optimal combining matrix for the whitened signal $\A_{kc}^{-1/2}\y_k$ is $\A_{kc}^{-1/2}\overline{\G}_{kc} \overline{\W}_{kc}$. 
Therefore, the desired part of the signal at \gls{ue} $k$ from cluster $c$ is 
\begin{align}
    \nonumber
    (\A_{kc}^{-\frac{1}{2}}\overline{\G}_{kc} \overline{\W}_{kc})^{\tH} \A_{kc}^{-\frac{1}{2}}\overline{\G}_{kc} \overline{\W}_{kc} \q_{kc}  
    \\
    = (\A_{kc}^{-1}\overline{\G}_{kc} \overline{\W}_{kc})^{\tH} \overline{\G}_{kc} \overline{\W}_{kc} \q_{kc}, 
\end{align}
which means that the optimal combining matrix is $\widetilde{\V}_{kc}$. 
	
Using the matrix inversion lemma and after some matrix manipulations~\cite{bjornson2017massive}, the relationship between the optimal combining $\widetilde{\V}_{kc}$ and the \gls{mmse} combining matrix $\overline{\V}_{kc} = \overline{\V}_{kc}^\text{MMSE}$ in \eqref{eqn:Combiner_MMSE_exp} can be found. 
Consider the term 
\begin{align}
    \nonumber
    \A_{kc}^{-1}\overline{\G}_{kc} \overline{\W}_{kc}
    & = \Brac{\A_{kc} + \overline{\G}_{kc} \overline{\W}_{kc} \overline{\W}_{kc}^{\tH} \overline{\G}_{kc}^{\tH}}^{-1} \overline{\G}_{kc} \overline{\W}_{kc} 
    \\ 
    & \times \Brac{\II_{d_{kc}} \!+ \overline{\W}_{kc} \overline{\G}_{kc} \A_{kc}^{-1} \overline{\G}_{kc} \overline{\W}_{kc}}^{-1}.
\end{align}
Letting $\B_{kc} = (\II_{d_{kc}} + \overline{\W}_{kc}^{\tH} \overline{\G}_{kc}^{\tH} \A_{kc}^{-1} \overline{\G}_{kc} \overline{\W}_{kc} )^{-1}$. It is true that $\widetilde{\V}_{kc} = \overline{\V}_{kc}^\text{MMSE}\B_{kc}$.
Also, the maximum instantaneous rate $R_{kc}$ obtained using $\widetilde{\V}_{kc}$ is
\begin{align}
    \label{eqn:OptimalRate}
    \nonumber
    R_{kc} (\widetilde{\V}_{kc})  & = \log_2 \frac{\Abs{\widetilde{\V}_{kc}^{\tH}\Brac{\A_{kc} + \overline{\G}_{kc} \overline{\W}_{kc} \overline{\W}_{kc}^{\tH} \overline{\G}_{kc}^{\tH}}\widetilde{\V}_{kc}}}{\Abs{\widetilde{\V}_{kc}^{\tH}\A_{kc}\widetilde{\V}_{kc}}}  
    \\
    \nonumber
    & = \log_2 \frac{\Abs{\B_{kc}^{\tH}\!\overline{\V}_{kc}^{\tH}\!\! \Brac{ \!\!\A_{kc} \!\! + \overline{\G}_{kc} \overline{\W}_{kc} \overline{\W}_{kc}^{\tH} \overline{\G}_{kc}^{\tH}\!}\!\!\overline{\V}_{kc}\!\B_{kc}}}{\Abs{\B_{kc}^{\tH}\overline{\V}_{kc}^{\tH}\A_{kc}\overline{\V}_{kc}\B_{kc}}} 
    \\
    \nonumber
    & = \log_2 \frac{\Abs{\overline{\V}_{kc}^{\tH}\!\! \Brac{ \!\!\A_{kc} \!\! + \overline{\G}_{kc} \overline{\W}_{kc} \overline{\W}_{kc}^{\tH} \overline{\G}_{kc}^{\tH}\!}\!\!\overline{\V}_{kc}}}{\Abs{\overline{\V}_{kc}^{\tH}\A_{kc}\overline{\V}_{kc}}} 
    \\
    & = R_{kc} (\overline{\V}_{kc}) = R_{kc} (\overline{\V}_{kc}^\text{MMSE} ).
\end{align}
The third equality in \eqref{eqn:OptimalRate} holds due to the fact that $\Abs{\X\Y} = \Abs{\X}\Abs{\Y}$ if $\X$ and $\Y$ are square matrices. 
Hence, from \eqref{eqn:OptimalRate}, $\overline{\V}_{kc}^\text{MMSE} $ can achieve the maximum instantaneous rate $R_{kc}$. 
Substituting $\eqref{eqn:Combiner_MMSE_exp}$ in \eqref{eqn:PCNC_Rkc} gives \eqref{eqn:RateWith_MMSE_exp}, which completes the proof.

\bibliographystyle{IEEEtran}
\bibliography{IEEEabrv,references}

\begin{IEEEbiography}[{\includegraphics[width=1in,height=1.25in,clip,keepaspectratio]{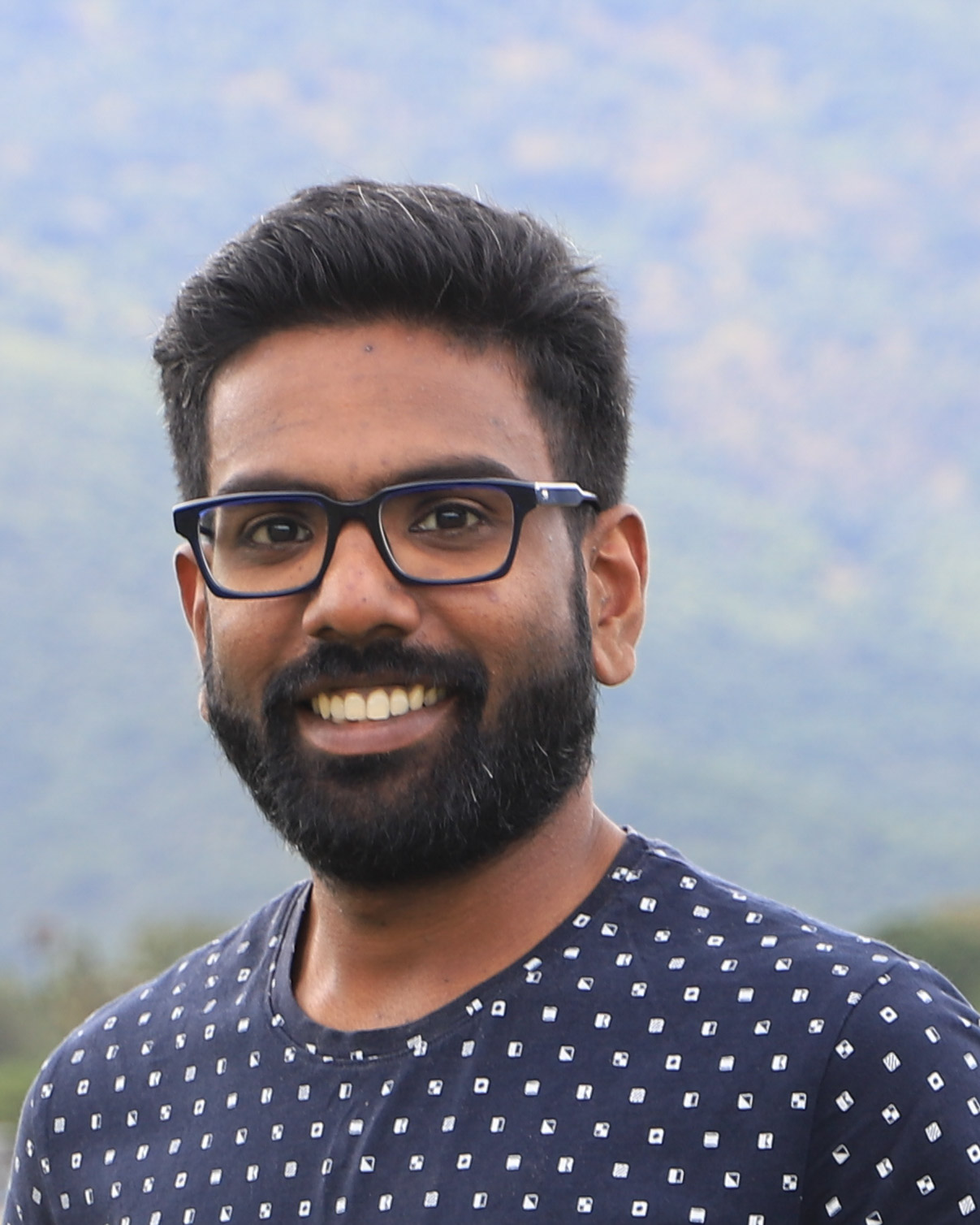}}]{Unnikrishnan Kunnath Ganesan}~(Graduate Student Member) received the Bachelor of Technology degree in Electronics and Communication Engineering from University of Calicut, Kerala, India in 2011 and the Masters in Engineering degree in Telecommunication Engineering from Indian Institute of Science, Bangalore, India in 2014. 
	From 2014 to 2017, he worked as modem systems engineer with Qualcomm India Private Limited, Bangalore and from 2017 to 2019 he worked as senior firmware engineer with Intel. 
	He is currently pursuing the Ph.D. degree with the Department of Electrical Engineering (ISY), Link\"oping University, Sweden. 
	His primary research interests includes MIMO wireless communications, signal processing, and information theory.
\end{IEEEbiography}

\begin{IEEEbiography}[{\includegraphics[width=1in,height=1.25in,clip,keepaspectratio]{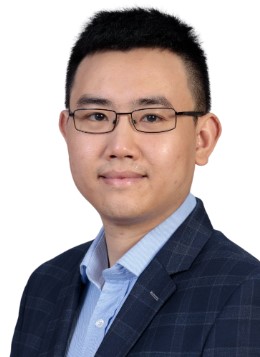}}]
	{Tung Thanh Vu}~(Member, IEEE)~received a Ph.D. degree in wireless communications from The University of Newcastle, Australia, in 2021. He is currently a Research Fellow at the School of Engineering, Macquarie University, Australia. His research interests include optimization, communication theories, and machine learning applications for 5G-and-beyond wireless networks, especially with massive MIMO, cell-free massive MIMO, federated learning, full-duplex communications, physical layer security, and low-earth orbit satellite communications.
	
	Dr. Tung Thanh Vu is serving as an Editor of Elsevier \textit{Physical Communication} (PHYCOM). He has also served as a member of the technical program committee and the symposium/session chair at several IEEE international conferences such as GLOBECOM, ICCE, and ATC. He was an \textsc{IEEE Wireless Communications Letters} exemplary reviewer for 2020 and 2021 and an \textsc{IEEE Transactions on Communications} exemplary reviewer for 2021. He received the Best Poster Award at the AMSI Optimise Conference in 2018. 
\end{IEEEbiography}

\begin{IEEEbiography}[{\includegraphics[width=1in,height=1.25in,clip,keepaspectratio]{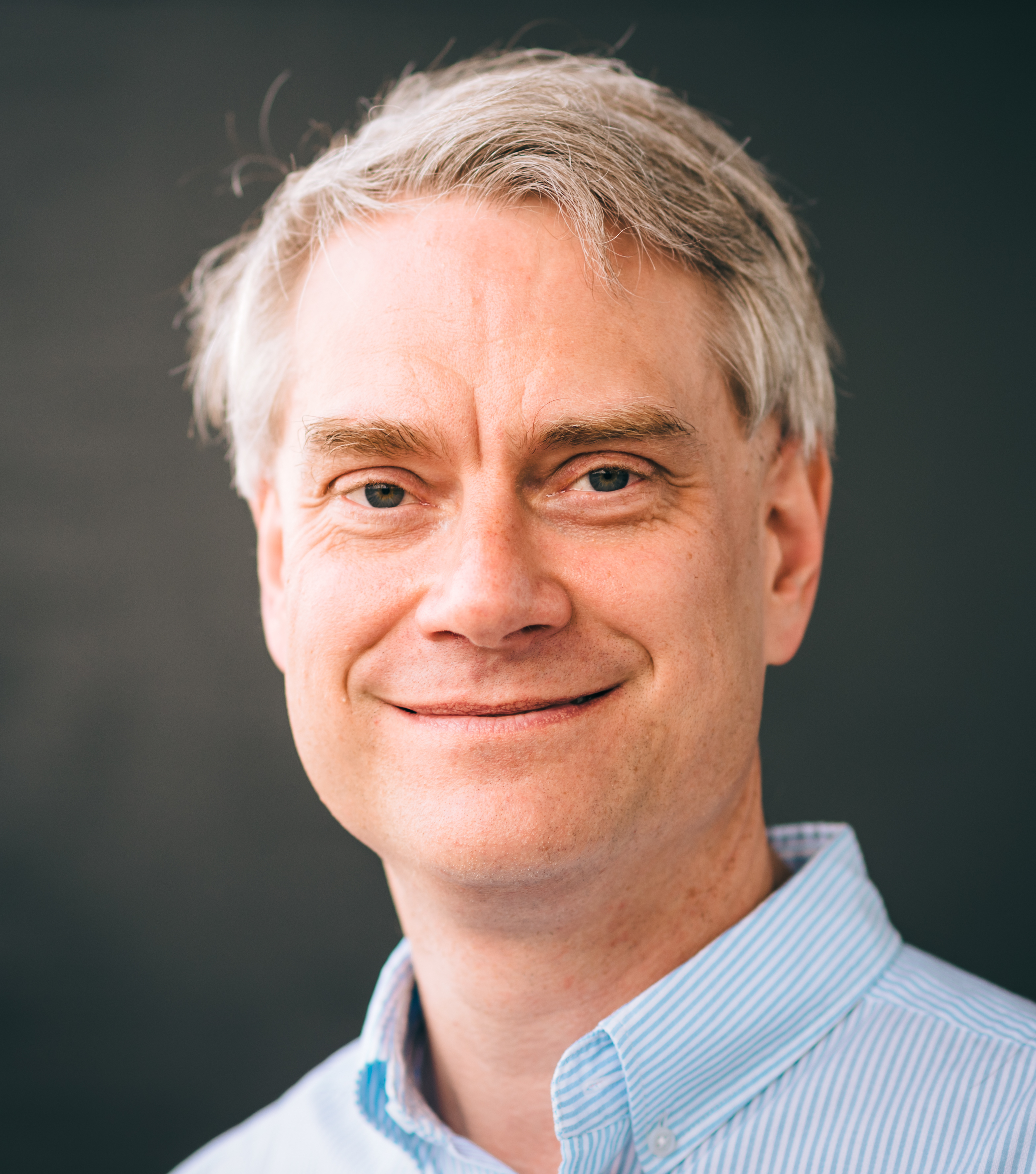}}]{Erik G. Larsson}~~(Fellow, IEEE) received the Ph.D. degree from Uppsala University,
	Uppsala, Sweden, in 2002.  He is currently Professor of Communication
	Systems at Link\"oping University (LiU) in Link\"oping, Sweden. He was
	with the KTH Royal Institute of Technology in Stockholm, Sweden, the
	George Washington University, USA, the University of Florida, USA, and
	Ericsson Research, Sweden.  His main professional interests are within
	the areas of wireless communications and signal processing. He 
	co-authored \emph{Space-Time Block Coding for  Wireless Communications} (Cambridge University Press, 2003) 
	and \emph{Fundamentals of Massive MIMO} (Cambridge University Press, 2016).

	He served as  chair  of the IEEE Signal Processing Society SPCOM technical committee (2015--2016), 
	chair of  the \emph{IEEE Wireless  Communications Letters} steering committee (2014--2015), 
	member of the  \emph{IEEE Transactions on Wireless Communications}    steering committee (2019-2022),
	General and Technical Chair of the Asilomar SSC conference (2015, 2012), 
	technical co-chair of the IEEE Communication Theory Workshop (2019), 
	and   member of the  IEEE Signal Processing Society Awards Board (2017--2019).
	He was Associate Editor for, among others, the \emph{IEEE Transactions on Communications} (2010-2014), 
	the \emph{IEEE Transactions on Signal Processing} (2006-2010),
	and  the \emph{IEEE Signal  Processing Magazine} (2018-2022).
	
	He received the IEEE Signal Processing Magazine Best Column Award
	twice, in 2012 and 2014, the IEEE ComSoc Stephen O. Rice Prize in
	Communications Theory in 2015, the IEEE ComSoc Leonard G. Abraham
	Prize in 2017, the IEEE ComSoc Best Tutorial Paper Award in 2018, the
	IEEE ComSoc Fred W. Ellersick Prize in 2019, and the IEEE SPS Donald
	G. Fink Overview Paper Award in 2023.  He is a member of the Swedish
	Royal Academy of Sciences (KVA), and Highly Cited according to ISI Web
	of Science.
\end{IEEEbiography}

\end{document}